\documentclass[preprint,amsmath,amssymb,aps,amsart,floatfix]{article}
\usepackage[title]{appendix}

\usepackage{graphics}
\usepackage{amsmath,amssymb,amsfonts}
\usepackage{mathtools}
\usepackage{hyperref}

\newcommand {\bea}{\begin{eqnarray}}
\newcommand {\eea}{\end{eqnarray}}
\newcommand {\be}{\begin{equation}}
\newcommand {\ee}{\end{equation}}

\newcommand {\non}{\nonumber\\}
\newcommand {\nonu}{\nonumber}
\newcommand {\ub}{\underbrace}
\newcommand {\dss}{\displaystyle\sum}
\newcommand {\ob}{\overbracket}
\newcommand {\dg}{{\dagger}}
\newcommand {\adag}{{a^\dagger}}

\usepackage{url}
\usepackage{mathtools}
\usepackage{authblk}

\DeclarePairedDelimiter\bra{\langle}{\rvert}
\DeclarePairedDelimiter\ket{\lvert}{\rangle}
\DeclarePairedDelimiterX\braket[2]{\langle}{\rangle}{#1
\delimsize\vert #2}

\usepackage[bottom]{footmisc}

\begin{document}

\title{Pairing, Quasi-spin and Seniority }

\author[1,2, \footnote{Email: bijay.agrawal@saha.ac.in (Bijay Agrawal)}]{Bijay Kumar Agrawal }

\affil[1]{ Saha Institute of Nuclear Physics, Kolkata, India. }
\affil[2]{Homi Bhabha National Institute, Mumbai, India.}

\author[3]{Bhoomika Maheshwari}
\affil[3]{ University of Malaya, Kuala Lumpur, Malaysia.  }

\date{\today}


%


\maketitle

\begin{abstract}
We present our concise notes for the lectures and tutorials  on
pairing, quasi-spin and seniority delivered  at SERB school on Role of
Symmetries in Nuclear Physics, AMITY University, 2019. Starting with some
generic features of residual nucleon-nucleon interactions, we provide
detailed derivation of the matrix elements for the $\delta$-interaction
which is the basis for the standard pairing Hamiltonian.  The eigen
values for  standard pairing Hamiltonian are then obtained within the
quasi-spin formalism. The algebra involving quasi-spin operators is
performed explicitly using the annihilation and creation operators
for single nucleon together with the application of Wick's theorem.
These techniques are expected to be helpful in deriving  the mean-field
equations for the Hartree-Fock, Bardeen–Cooper–Schrieffer  and
Hartree-Fock Bogoliubov theories.  
\end{abstract}

\section{Introduction}
\label{intro}
Role of symmetries is important when no exact solutions to a physical
problem are known, such as nuclear force. More often, our problem
is analogous to the atomic and molecular structure. However, things
are simpler there, as Coulomb forces are well known. In case of the
nuclear force, the knowledge of its mathematical form is still an
open question. Fortunately, due to symmetry properties of the basic
interactions in most physical systems, qualitative features of the
composite system are not too sensitive to the details of the interaction
itself. Symmetries of physical laws may lead to the laws of conservation
of  spin, isospin and energy. For example, the orbital, spin and isospin
quantum operators usually commute with the nuclear interaction Hamiltonian. This means the interaction Hamiltonian commutes with all rotations in orbital,
spin and isospin spaces. Therefore, the angular momentum coupling and
spherical harmonics play a vital role in understanding various properties
of nuclei.

One of the most important inputs to the system of interacting nucleons
is the matrix elements for the appropriate nucleon-nucleon interaction.
These matrix elements can be evaluated by decomposing nucleon-nucleon
interaction in to its radial and angular parts.  The evaluation of
two-body matrix elements, thus, reduces to the calculation of the radial
and the angular matrix elements.  The procedure for the calculation of
the two-body matrix elements for different interactions mainly differ in
their radial matrix elements. The angular part usually depends on the
product of the spherical harmonics corresponding to the two nucleons.
The angular part of the matrix elements is straightforward but lengthy,
since, they are evaluated within the anti-symmetrized  coupled states
of two nucleons. Further more, each of the nucleonic wave function
correspond to the coupled state of the their orbital angular momentum
and intrinsic spin. The derivation is not available in full detail at a single place, which we found important to discuss with students during SERB school.

In the present article, we summarize details  of our lectures  on the
evaluation of the direct and exchange terms for the matrix elements
of the $\delta$-interaction.  Next, we consider the standard
pairing Hamiltonian which mimics the short-range nature of the
$\delta$-interaction. Its exact eigen values and eigen states are  obtained
within the quasi-spin  formalism.  The algebra involving quasi-spin
operators is performed directly in terms of the annihilation and creation
operators for single nucleon. We
simplify our  algebra using the Wick's theorem.  These techniques are
also handy  in deriving the mean-field equations corresponding to
different  trial wave-function which leads to the various mean-field
approaches such as Hartree-Fock, Bardeen-Cooper-Schrieffer and
Hartree-Fock Bogoliubov approximations.  Aim of the present notes is to
provide sufficient details in a self-contained manner. In Appendices,
we include various identities involving 3j- and 6j- symbols and provide
the derivations for the quasi-spin operators and the reduced matrix
elements for the spherical harmonics which were covered during tutorials.

\section
{Generic features of residual interaction}

The binding energy for several hundreds of nuclei are known very
precisely. The variations of binding energy per nucleon with mass
number clearly manifests the strong and short range nature of the
nuclear force. The Hamiltonian for such an strongly interacting system
is not exactly solvable. A general nuclear Hamiltonian satisfy

\begin{eqnarray}
H \Psi (r_1, r_2, r_3,... r_A) = E \Psi (r_1, r_2, r_3,... r_A)
\end{eqnarray} 
where $H$ comprises the kinetic energy and interaction part,
\begin{eqnarray}
H = \sum_{i=1}^A T(r_i) + \sum_{i < j}^A V (r_{ij}) + \sum_{i < j < k}^A V (r_{ijk})+....
\end{eqnarray} 
where $T(r_i)$ is kinetic energy, $V (r_{ij})$ and $V (r_{ijk})$ are two-body and three-body parts of interaction. The exact solution to such nuclear Hamiltonian require a diagonalization of Hamiltonian matrix whose dimensions are infinitely large. Usually, one decomposes the total Hamiltonian
into the that for the system of non-interacting nucleons moving in an
average potential and the left over part termed as residual
interaction. The average potential is, to a good approximation, taken to be the harmonic oscillator potential or the one obtained by averaging the
nucleon-nucleon interactions. It may be emphasized that the minor
perturbation created by residual interaction is so important that it
should not be ignored.  In fact, these residual interactions determine
almost everything we know about most of the nuclei.

\begin{figure}
\begin{center}
\includegraphics[width=1.0\columnwidth]{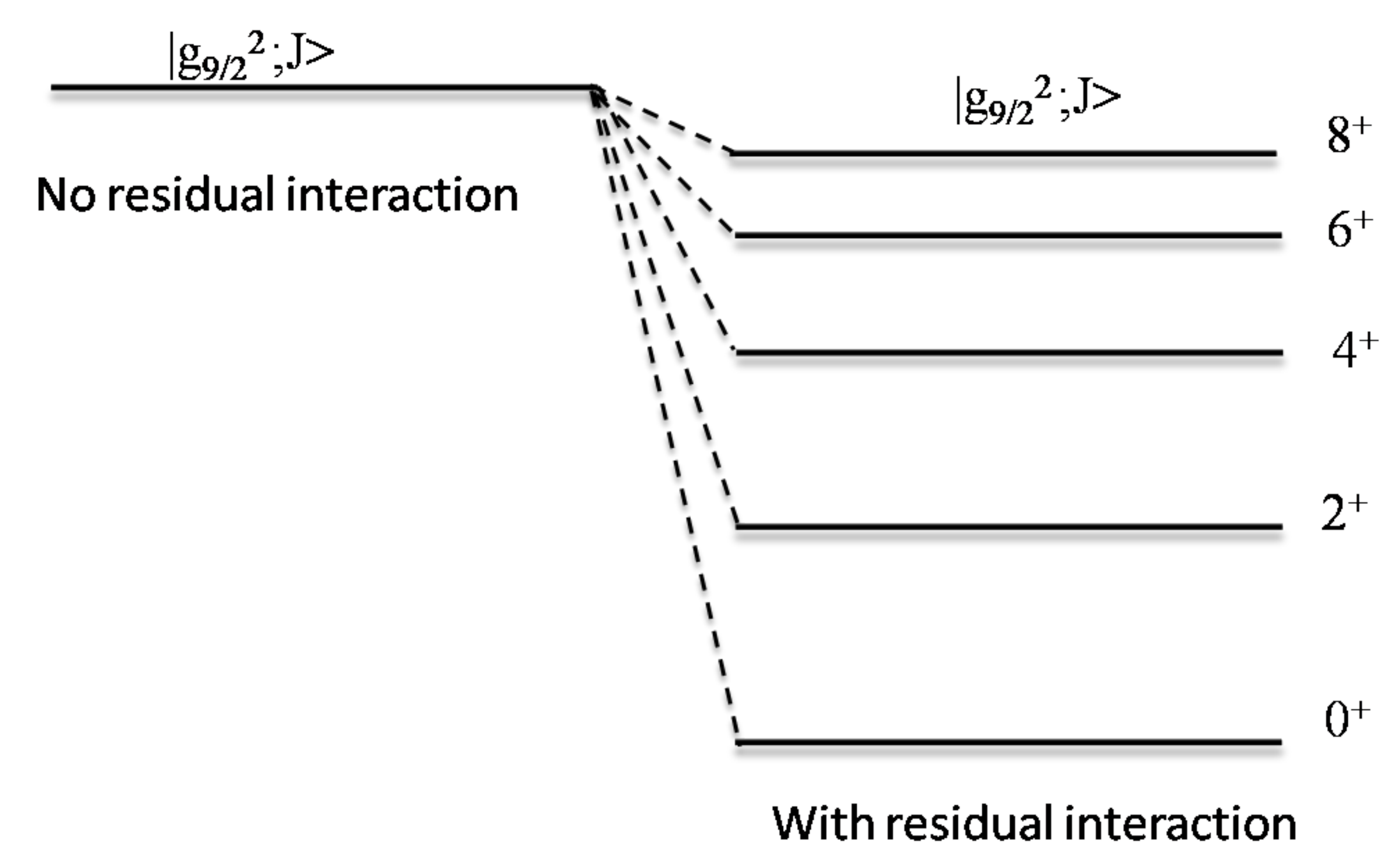}
\caption{\label{example} A schematic diagram showing the energy levels with
and without residual interaction for $g_{9/2}^2$ configuration.}
\end{center}
\end{figure}

The importance of residual interactions can be easily realized
through the energy spectra one obtains for two-particles system.  Let us
consider two identical nucleons in $g_{9/2}^2$ and interacting through a residual
interaction $V_{12}$. The allowed values for the total angular momentum
$J$ ranges from $ |j_1-j_2 | $ to $j_1+j_2$ (triangular inequalities),
where, $ j_1, j_2$ are the angular momentum for the single nucleons.
The allowed values for $J$ would be $0,2,4,6,8$ for $g_{9/2}^2$
configuration. The resulting $J$ depends on the angle between two orbital
planes. A schematic energy level scheme for these two nucleons is shown
in Fig. ~\ref{example}, with or without residual interaction. Without
residual interaction, all the $J$ states would be degenerate while
the states with the inclusion of the  residual interaction would
be non-degenerate producing significant energy splitting.  Residual
interaction $V_{12}$ is the strongest when two particles are closest
to each other, i.e. when orbitals are co-planar. So, the strongest
interaction takes place either  for $J_{min} = | j_1 - j_2 | =0 $, or $J_{max} =
j_1 + j_2 = 8 $. According to the Pauli principle, no two fermions can
occupy the same state/place. Total wave functions must be anti-symmetric.

The  wave functions for nucleons are characterized  by the spatial
co-ordinates and spins.  The spatial part of two-nucleon wave function
can be decomposed  into its center of mass and relative coordinates. The
anti-symmetrization of the spatial part of wave-function requires,

\begin{eqnarray}
\Psi(-\vec{r})=-\Psi(\vec{r}) \\
\text{where} \quad \vec{r}=\vec{r_2}-\vec{r_1}. \nonumber
\end{eqnarray}
This ensures that the anti-symmetrized spatial part of wave-function
vanishes, if both particles are at the same place, i.e.,  $\Psi(0)=0 $.
On dividing the total wave functions into spatial and spin parts we get,
\begin{eqnarray}
\Psi_{total}=\Psi_{spatial} (L) \times \Psi_{spin} (S) = Anti-symmetric 
\end{eqnarray}
So, there will be two possibilities for $\Psi_{total}$ to be anti-symmetric:\\
(a) If $\Psi_{spatial}$ is anti-symmetric then $\Psi_{spin}$ would be symmetric.\\
(b) If $\Psi_{spatial}$ is symmetric then $\Psi_{spin}$ would be anti-symmetric. Also, \\
\begin{eqnarray}
S=1/2+1/2=1 \longrightarrow  \Psi_{spin} \quad symmetric, \nonumber \\ 
S=1/2-1/2=0 \longrightarrow  \Psi_{spin} \quad anti-symmetric. \nonumber
\end{eqnarray}

Let us consider a simple form of residual interaction, to be
a $\delta$-interaction. The $\delta$-interaction acts only when
$\vec{r}=0$, requires $\Psi_{spatial}$ to be symmetric. Thus $\delta$
-interaction  acts on two identical nucleons with  $S=0$. As a result,
the $S=0$ states is lower in energy. For two nucleons in $g_{9/2}^2$,
the allowed $J_{min}=0$ would belong to the $S=0$ while $J_{max}=8$
would belong to the $S=1$. This means that $J=0$ is the lowest since
$S=0$ state  is lowered due to the action of $\delta$-interaction as
shown schematically  in Fig.~\ref{example}. These general features
of the residual interaction can be easily understood in terms of the
two-body matrix elements for the  $\delta$-interaction which we cover
in the following section.

\section{Matrix elements of $\delta$-interaction}

The simplest interaction potential such as the $\delta$- and Yukawa forces
depends only on the distance between two nucleons. Two-body matrix
elements of these interactions are evaluated either by decomposing
the two-nucleon wave functions into their center of mass and relative
co-ordinates or directly in terms of the two nucleon wave functions
which are the anti-symmetrized products or sum of the anti-symmetrized
products of a single-nucleon wave function. In the later case, one needs
to decompose the interaction potentials into radial and angular parts
using the multipole expansion as follows.

\subsection{Multipole expansion}

The interaction
$V(\mathbf{r}_1,\mathbf{r}_2) \equiv V(| \mathbf{r}_1 -\mathbf{r}_2|)$
can be expanded in terms of  various multipoles as \cite{Talmi63,Talmi93},

\bea
V(|\mathbf{r}_1 -\mathbf{r}_2|) = \sum_{K=0}^\infty V_K(r_1, r_2) P_K(cos\omega_{12}),
\eea
where, $P_K(cos\omega_{12})$ is the Legendre  polynomial and
$ \omega_{12} =( \mathbf{r}_1 \cdot \mathbf{r}_2)/r_1 r_2$. The
function $V_K(r_1, r_2)$ are symmetric in $r_1$ and $ r_2$.  Due to the
orthogonality relations of Legendre polynomials, the $V_K(r_1, r_2)$
may be obtained from $V(|\mathbf{r}_1 -\mathbf{r}_2|)$ as,

\bea
V_K(r_1, r_2) =  \frac{2K+1}{2}\int_{-1}^{+1} V(|\mathbf{r}_1 -\mathbf{r}_2|)P_K(cos\omega_{12}) d cos\omega_{12}.
\eea
The Legendre polynomials in the above equation can be expressed by the
addition theorem of spherical harmonics as,

\bea
P_K(cos\omega_{12}) =\frac{4\pi}{2K+1}
\sum_{q} (-1)^q Y^K_q
(\theta_1, \phi_1)  Y^K_{-q} (\theta_2, \phi_2). 
\eea

In general,  the radial part of the matrix elements is evaluated
numerically and  it is   different for different interactions. However,
the angular part of the matrix elements is evaluated analytically and the
the procedure to calculate it up to the large extent is similar for most
of the interactions. 

\subsection{ $\delta$-interaction}

We now expand the $\delta$-interaction in terms of the multipoles and evaluate its matrix elements between the anti-symmetrized wave functions for two nucleons.
The $\delta$-interaction between two nucleons is \cite{Talmi93}, 

\be
\delta(\vec{r})= \delta(\vec{r}_1-\vec{r}_2)\label{del_1}
\ee
with $\vec{r} = \vec{r}_1 - \vec{r}_2$, $\vec{r}_1$
and $\vec{r}_2$ being the coordinates of the two nucleons. The
Eq. (\ref{del_1}) can be expanded in terms of the multipoles as,

\bea
\delta(\vec{r}) &=& \sum_{K}\frac{\delta(r_1 - r_2)}{r_1 r_2}
 \frac{2K+1}{4\pi}P_K (cos\omega_{12}) \qquad
\text{(multipole decomposition)}\\
&=&\sum_{K}\frac{\delta(r_1 - r_2)}{r_1 r_2} \sum_{q} (-1)^q Y^K_q
(\theta_1, \phi_1)  Y^K_{-q} (\theta_2, \phi_2) \\
&=&\sum_{K}\frac{\delta(r_1 - r_2)}{r_1 r_2}  Y^K (1) \cdot  Y^K(2),
\label{del_2} 
\eea
where, $ Y^K (i)= Y^K (\theta_i, \phi_i)$.

The matrix elements of the interaction must be calculated within the anti-symmetrized states for two nucleons whose angular momentums are either
coupled or uncoupled. They are  usually evaluated within the  coupled
states of nucleons which are also anti-symmetrized.  The anti-symmetrized
two-nucleon coupled states can be expressed as,
\bea
\ket{\alpha_1 j_1 \alpha_2 j_2;JM \, TM_T}_{\bf{as}} &=&  
\frac{1}{\sqrt{2(1+\delta_{j_1 j_2} \delta_{t_1 t_2} )}} 
[\ket{\alpha_1 j_1 \alpha_2 j_2;JM}\non
&&- (-1)^{j_1+j_2 +J} (-1)^{t_1+ t_2+T}\ket{\alpha_2 j_2 \alpha_1 j_1;JM}
  ] \non
 \label{as_1}
\eea
where, $\alpha_i = n_i,l_i$ denotes the principle and orbital quantum
numbers, $j_i$ and  $t_i$ are the angular momentum and isospin of $i$-th
nucleon, respectively. The coupled angular momentum (isospin) and its
$z$-components are denoted by $J$($T$) and $M$($M_T$), respectively.
Since, the isospin $t_1 = t_2 = \frac{1}{2}$ for two nucleons and the interaction
considered Eq. (\ref{del_1}) is independent of isospin, the matrix
element for the interaction would depend on the total isospin $T$  only
through the phase factor which  ensures  the total anti-symmetrization
of the two-nucleon wave functions. Hence forth, we retain  the isospin
quantum number $T$  only in the phase factor.  Further, the
$\delta$-interaction conserves the total angular momentum due to its
rotational invariance, the  matrix elements are non-zero only if the
total angular momentum and its $z-$component  in the initial and final
states are same. Matrix element of the $\delta$-interaction between
the anti-symmetrized wave function for two nucleons can be written as,

\bea
\bra{\alpha_1 j_1 \alpha_2 j_2;JM} \delta(\vec{r}) \ket{\alpha'_1 j'_1
\alpha'_2 j'_2;JM} _{\bf{as}}
&=&N_{12}
 [\ub{ \bra{\alpha_1 j_1 \alpha_2 j_2;JM} \delta(\vec{r})
\ket{\alpha'_1 j'_1 \alpha'_2 j'_2;JM} }_{direct}\non
&+&\ub{(-1)^{j'_1+j'_2 +J+T} \bra{\alpha_1 j_1 \alpha_2 j_2;JM}
\delta(\vec{r}) \ket{\alpha'_2 j'_2
\alpha'_1 j'_1;JM}}_{exchange}]  \label{mat_1} \non
\\
\text{with}\qquad  N_{12}& =& [(1+\delta_{j_1 j_2})(1+\delta_{j'_1 j'_2})]^{-1/2} . \label{n12}
\eea

\subsubsection{Direct term}
By substituting Eq.(\ref{del_2}) in Eq.(\ref{mat_1}),  the direct term becomes,

\bea
D&=&  \bra{\alpha_1 j_1 \alpha_2 j_2;JM} 
\sum_{K} \frac{\delta(r_1 - r_2)}{r_1 r_2}
  Y^K (1) \cdot  Y^K(2) \ket{\alpha'_1 j'_1 \alpha'_2 j'_2;JM}  \label{mat_2}
\eea
which can be decomposed into the radial and the angular parts. The radial
part consists of the matrix element for $\delta(r_1 - r_2)/r_1 r_2$ and the
angular part corresponds to the matrix element of $Y_K(1)\cdot Y_K(2)$. The
angular part can be further decomposed into the products of matrix
elements for $Y_K(1)$ and $Y_K(2)$ using Eq. (\ref{yk12}). The
Eq. (\ref{mat_2}) simplifies to,

\bea
D&=& {4\pi}
C_0 \sum_{K} (-1)^{j_2+J+j'_1}
\left \{ \begin{matrix}
j_1&j_2&J\\
j'_2&j'_1&K\\
\end{matrix} \right \} \non
&&(l_1\frac{1}{2}; j_1 ||Y^K|| l'_1 \frac{1}{2};j'_1)
(l_2\frac{1}{2}; j_2 ||Y^K|| l'_2 \frac{1}{2};j'_2)  \label{direct}
\eea
where $C_0$ denotes the radial integral as follows:
\bea
C_0&=&\frac{1}{4\pi}\int R^*_{n_1 l_1 j_1}(r_1)R^*_{n_2 l_2 j_2}(r_2)
\frac{\delta(r_1 - r_2)}{r_1 r_2}
R_{n'_1 l'_1 j'_1}(r_1)R_{n'_2 l'_2 j'_2}(r_2)r_1^2 r_2^2
dr_1 dr_2\non
&=&\frac{1}{4\pi}\int R^*_{n_1 l_1 j_1}(r)R^*_{n_2 l_2 j_2}(r)
R_{n'_1 l'_1 j'_1}(r)R_{n'_2 l'_2 j'_2}(r)r^2 dr \label{c0}
\eea
The reduced matrix elements appearing in Eq. (\ref{direct}) can be simplified
using Eq.(\ref{yk1_yk2}) as, 

\bea
(l_1\frac{1}{2}; j_1 ||Y^K|| l'_1 \frac{1}{2};j'_1)
&=& (-1)^{j_1 -\frac{1}{2}}
\left( \frac{\hat{j'_1}\hat{j_1}\hat{K}}{\sqrt{4\pi}}\right)  \nonumber \\
 &&\frac{1}{2}(1+(-1)^{l'_1+l_1+K})
\times \left (
\begin{matrix}
 j_1& K&j'_1\\
-\frac{1}{2}& 0&\frac{1}{2}
\end{matrix}
\right ) \\
(l_1\frac{1}{2}; j_1 ||Y^K|| l'_1 \frac{1}{2};j'_1)
(l_2\frac{1}{2}; j_2 ||Y^K|| l'_2 \frac{1}{2};j'_2) 
&=& (-1)^{j_1+j_2 -1}
\left( \frac{\hat{j'_1}\hat{j_1}\hat{j'_2}\hat{j_2}
\hat{K}^2}{4\pi}\right)
 \nonumber  \\
 &&\frac{1}{4}(1+(-1)^{l'_1+l_1+K}) (1+(-1)^{l'_2+l_2+K})\non
&&\times \left (
\begin{matrix}
 j_1& K&j'_1\\
-\frac{1}{2}& 0&\frac{1}{2}
\end{matrix}
\right )   \left (
\begin{matrix}
 j_2& K&j'_2\\
-\frac{1}{2}& 0&\frac{1}{2}
\end{matrix}
\right ) \label{rme}
\eea
In the Eq.(\ref{rme}) $\hat{j}_i=\sqrt{2j_i+1}$, $\hat{K}=\sqrt{2K+1}$.
For simplicity, let us consider $ l_i',j_i'= l_i,j_i$. So Eq. (\ref{rme}) becomes,

\bea
(l_1\frac{1}{2}; j_1 ||Y^K|| l_1 \frac{1}{2};j_1)
(l_2\frac{1}{2}; j_2 ||Y^K|| l_2 \frac{1}{2};j_2) 
&=& (-1)^{j_1+j_2 -1}
\left( \frac{\hat{j_1}^2\hat{j_2}^2 \hat{K}^2}{4\pi}\right)\non
&&\times \left (
\begin{matrix}
 j_1& K&j_1\\
-\frac{1}{2}& 0&\frac{1}{2}
\end{matrix}
\right )   \left (
\begin{matrix}
 j_2& K&j_2\\
-\frac{1}{2}& 0&\frac{1}{2}
\end{matrix}
\right )\non
&& \frac{1}{4}(1+(-1)^{2l_1+K}) (1+(-1)^{2l_2+K})\non
&=& (-1)^{j_1+j_2 -1}
\left( \frac{\hat{j_1}^2\hat{j_2}^2 \hat{K}^2}{4\pi}\right)
\frac{1}{2}(1+(-1)^K) \non
&&\times \left (
\begin{matrix}
 j_1& K&j_1\\
-\frac{1}{2}& 0&\frac{1}{2}
\end{matrix}
\right )   \left (
\begin{matrix}
 j_2& K&j_2\\
-\frac{1}{2}& 0&\frac{1}{2}
\end{matrix}
\right )\label{TME}
\eea
Substituting Eq. (\ref{TME}) in Eq. (\ref{direct}) we get,
\bea
D &=& 4 \pi C_0 \sum_{K} (-1)^{j_2+J+j_1}
\left \{ \begin{matrix}
j_1&j_2&J\\
j_2&j_1&K\\
\end{matrix} \right \} 
(-1)^{j_1+j_2 -1}
\left( \frac{\hat{j_1}^2\hat{j_2}^2 \hat{K}^2}{4\pi}\right)
 \nonumber  \\
 && \frac{1}{2}(1+(-1)^K)
\times \left (
\begin{matrix}
 j_1& K&j_1\\
-\frac{1}{2}& 0&\frac{1}{2}
\end{matrix}
\right )   \left (
\begin{matrix}
 j_2& K&j_2\\
-\frac{1}{2}& 0&\frac{1}{2}
\end{matrix}
\right )\non
&=& C_0 \left( \frac{\hat{j_1}^2\hat{j_2}^2 }{2}\right)
\sum_{K} (-1)^{J-1}\hat{K}^2 (1+(-1)^K)
\times \left \{ \begin{matrix}
j_1&j_2&J\\
j_2&j_1&K\\
\end{matrix} \right \}
 \left (
\begin{matrix}
 j_1& K&j_1\\
-\frac{1}{2}& 0&\frac{1}{2}
\end{matrix}
\right )   \left (
\begin{matrix}
 j_2& K&j_2\\
-\frac{1}{2}& 0&\frac{1}{2}
\end{matrix}
\right )\label{TME17}\non
&=& C_0 \left( \frac{\hat{j_1}^2\hat{j_2}^2 }{2}\right)
\sum_{K}\left [ (-1)^{J-1}\hat{K}^2
\times \left \{ \begin{matrix}
j_1&j_2&J\\
j_2&j_1&K\\
\end{matrix} \right \}
 \left (
\begin{matrix}
 j_1& K&j_1\\
-\frac{1}{2}& 0&\frac{1}{2}
\end{matrix}
\right )   \left (
\begin{matrix}
 j_2& K&j_2\\
-\frac{1}{2}& 0&\frac{1}{2}
\end{matrix}
\right ) \right .\non
&& +(-1)^{J+K-1}\hat{K}^2
\left . \times \left \{ \begin{matrix}
j_1&j_2&J\\
j_2&j_1&K\\
\end{matrix} \right \}
 \left (
\begin{matrix}
 j_1& K&j_1\\
-\frac{1}{2}& 0&\frac{1}{2}
\end{matrix}
\right )   \left (
\begin{matrix}
 j_2& K&j_2\\
-\frac{1}{2}& 0&\frac{1}{2}
\end{matrix}
\right ) \right] \label{k-sum}
\eea
The sum over $K$ in Eq.(\ref{k-sum}) can be performed analytically.  
It  can be reduced to product of two 3j- symbols using a suitable identity as follows,

\bea
\dss_{M_3} \left (
\begin{matrix}
j_1 &j_2& J_3\\
m_1& m_2& M_3\\
\end{matrix}
\right )  \left (
\begin{matrix}
j'_1 &j'_2& J_3\\
m'_1& m'_2& -M_3\\
\end{matrix}
\right )
&=&\dss_{J'_3M'_3}(-1)^{j_3+j'_3+ m_1+ m'_1} \hat{J'_3}^2
\left \{
\begin{matrix}
j_1&j_2&J_3\\
j'_1&j'_2&J'_3\\
\end{matrix}
\right \}  \non
&& \left (
\begin{matrix}
j'_1 &j_2& J'_3\\
m'_1& m_2& M'_3\\
\end{matrix}
\right )
\left (
\begin{matrix}
j_1 &j'_2& J'_3\\
m_1& m'_2& -M'_3\\
\end{matrix}
\right )\label{TM24}
\eea
Substituting $J_3 = J$, $J_3' = K$, $j'_1=j_2$ and  $j'_2=j_1$ in Eq.
(\ref{TM24}) we get,

\bea
\dss_{M_3} \left (
\begin{matrix}
j_1 &j_2& J\\
m_1& m_2& M_3\\
\end{matrix}
\right )  \left (
\begin{matrix}
j_2 &j_1& J\\
m'_1& m'_2& -M_3\\
\end{matrix}
\right )
&=&\dss_{K M'_3}(-1)^{J+K+ m_1+ m'_1} \hat{K}^2
\left \{
\begin{matrix}
j_1&j_2&J\\
j_2&j_1&K\\
\end{matrix}
\right \}\non
&& \left (
\begin{matrix}
j_2 &j_2& K\\
m'_1& m_2& M'_3\\
\end{matrix}
\right )
\left (
\begin{matrix}
j_1 &j_1& K\\
m_1& m'_2& -M'_3\\
\end{matrix}
\right )\non
&=& \dss_{K M'_3}(-1)^{J+K+ m_1+ m'_1} \hat{K}^2
\left \{
\begin{matrix}
j_1&j_2&J\\
j_2&j_1&K\\
\end{matrix}
\right \}\non
&& (-1)^ {j_2 +j_2+K} \left (
\begin{matrix}
j_2 &K&j_2\\
m'_1& M'_3& m_2\\
\end{matrix}
\right )\non
&& (-1)^ {j_1 +j_1+K}
\left (
\begin{matrix}
j_1 & K&j_1\\
m_1& -M'_3& m'_2\\
\end{matrix}
\right )\non
&=& \dss_{K M'_3}(-1)^{J+K+ m_1+ m'_1} \hat{K}^2
\left \{
\begin{matrix}
j_1&j_2&J\\
j_2&j_1&K\\
\end{matrix}
\right \}\non
&& \left (
\begin{matrix}
j_2 &K&j_2\\
m'_1& M'_3& m_2\\
\end{matrix}
\right )
\left (
\begin{matrix}
j_1 & K&j_1\\
m_1& -M'_3& m'_2\\
\end{matrix}
\right )\label{TME20}
\eea
Sums over $M_3$ and $M'_3$ are constrained by,
\bea
m_1+m_2+M_3&=&  m'_1+m'_2 - M_3 = 0\non
m'_1+m_2+M'_3 &=& m_1+m'_2-M'_3= 0\nonu
\eea
Substituting $m_1 = m'_1 = -1/2 $, $m_2=m'_2= +1/2 $ in Eq.(\ref{TME20}), the
allowed values are $ M_3= M'_3=0 $\\
\bea
\left (
\begin{matrix}
j_1 &j_2& J\\
 - \frac{1}{2}& + \frac{1}{2}& 0\\
\end{matrix}
\right )  \left (
\begin{matrix}
j_2 &j_1& J\\
 - \frac{1}{2}& + \frac{1}{2}& 0\\
\end{matrix}
\right )
&=& \dss_{K }(-1)^{J+K-1} \hat{K}^2
\left \{
\begin{matrix}
j_1&j_2&J\\
j_2&j_1&K\\
\end{matrix}
\right \}\non
&& \left (
\begin{matrix}
j_2 &K&j_2\\
- \frac{1}{2}& 0& + \frac{1}{2}\\
\end{matrix}
\right )
\left (
\begin{matrix}
j_1 & K&j_1\\
- \frac{1}{2}& 0& + \frac{1}{2}\\
\end{matrix}
\right )\label{TME21}
\eea
Substituting $m_1 = m_2 = +1/2 $, $m'_1=m'_2= -1/2 $ and the allowed values of
$M_3= -1, M'_3=0$ in Eq.(\ref{TME20}) becomes,
\bea
\left (
\begin{matrix}
j_1 &j_2& J\\
  \frac{1}{2}& + \frac{1}{2}& -1\\
\end{matrix}
\right )  \left (
\begin{matrix}
j_2 &j_1& J\\
 - \frac{1}{2}& - \frac{1}{2}&1\\
\end{matrix}
\right )
&=& \dss_{K }(-1)^{J+K} \hat{K}^2
\left \{
\begin{matrix}
j_1&j_2&J\\
j_2&j_1&K\\
\end{matrix}
\right \}\non
&& \left (
\begin{matrix}
j_2 &K&j_2\\
- \frac{1}{2}& 0& + \frac{1}{2}\\
\end{matrix}
\right )
\left (
\begin{matrix}
j_1 & K&j_1\\
 \frac{1}{2}& 0& - \frac{1}{2}\\
\end{matrix}
\right )\label{TME22}\non
&=& \dss_{K }(-1)^{J+K} \hat{K}^2
\left \{
\begin{matrix}
j_1&j_2&J\\
j_2&j_1&K\\
\end{matrix}
\right \}\non
&& \left (
\begin{matrix}
j_2 &K&j_2\\
- \frac{1}{2}& 0& + \frac{1}{2}\\
\end{matrix}
\right ) (-1)^{2j_1 +K}
\left (
\begin{matrix}
j_1 & K&j_1\\
 -\frac{1}{2}& 0&  \frac{1}{2}\\
\end{matrix}
\right )\non
&=& \dss_{K }(-1)^{J-1} \hat{K}^2
\left \{
\begin{matrix}
j_1&j_2&J\\
j_2&j_1&K\\
\end{matrix}
\right \}\non
&& \left (
\begin{matrix}
j_2 &K&j_2\\
- \frac{1}{2}& 0& + \frac{1}{2}\\
\end{matrix}
\right )
\left (
\begin{matrix}
j_1 & K&j_1\\
 -\frac{1}{2}& 0&  \frac{1}{2}\\
\end{matrix}
\right )
\label{TME23}
\eea
Adding Eqs. (\ref{TME21}) and (\ref{TME23}), the R.H.S. and L.H.S. become,
\bea
RHS
&=&\dss_{K }(-1)^{J-1} (1+(-1)^K)\hat{K}^2
\left \{
\begin{matrix}
j_1&j_2&J\\
j_2&j_1&K\\
\end{matrix}
\right \}
\left (
\begin{matrix}
j_2 &K&j_2\\
- \frac{1}{2}& 0& + \frac{1}{2}\\
\end{matrix}
\right )
\left (
\begin{matrix}
j_1 & K&j_1\\
 -\frac{1}{2}& 0&  \frac{1}{2}\\
\end{matrix}
\right )\label{rhs}\\
LHS&=&
\left (
\begin{matrix}
j_1 &j_2& J\\
  \frac{1}{2}& + \frac{1}{2}& -1\\
\end{matrix}
\right )  \left (
\begin{matrix}
j_2 &j_1& J\\
 - \frac{1}{2}& - \frac{1}{2}&1\\
\end{matrix}
\right ) +
\left (
\begin{matrix}
j_1 &j_2& J\\
 - \frac{1}{2}& + \frac{1}{2}& 0\\
\end{matrix}
\right )  \left (
\begin{matrix}
j_2 &j_1& J\\
 - \frac{1}{2}& + \frac{1}{2}& 0\\
\end{matrix}
\right )\non
&&= \left (
\begin{matrix}
j_1 &j_2& J\\
  \frac{1}{2}& + \frac{1}{2}& -1\\
\end{matrix}
\right )^2 +\left (
\begin{matrix}
j_1 &j_2& J\\
 - \frac{1}{2}& + \frac{1}{2}& 0\\
\end{matrix}
\right )^2\label{TME24}
\eea
Using the recursion relation for 3j- symbol as in Eq. (\ref{rec1}), yields,
\bea
\left (
\begin{matrix}
j_1&j_2&J\\
 \frac{1}{2}& \frac{1}{2}&-1\\
\end{matrix}
\right )^2 &=&
\frac{[  \hat{j_1}^2 + \hat{j_2}^2 (-1)^{j_1+ j_2+J} ]^2}{4J(J+1)} 
 \left (
\begin{matrix}
j_1&j_2&J\\
-\frac{1}{2}&  \frac{1}{2}&0\\
\end{matrix}
\right )^2\non
LHS &=& \left[1 + \frac{[  \hat{j_1}^2 + \hat{j_2}^2 (-1)^{j_1+ j_2+J}
]^2}{4J(J+1)}\right]
 \left (
\begin{matrix}
j_1&j_2&J\\
-\frac{1}{2}&  \frac{1}{2}&0\\
\end{matrix}
\right )^2 \label{lhs}
\eea
Using Eqs. (\ref{k-sum}), (\ref{rhs}) and (\ref{lhs}), the direct term simply becomes,
\bea
D &=& C_0 \left( \frac{\hat{j_1}^2\hat{j_2}^2 }{2}\right)
 \times
 \left [1+ \frac{[\hat{j_1}^2 +
 \hat{j_2}^2 (-1)^{j_1+ j_2+J} ]^2}{4J(J+1)} \right
]\left (
\begin{matrix}
j_1&j_2&J\\
- \frac{1}{2}&  \frac{1}{2}&0\\
\end{matrix}
\right )^2 \label{direct_last}
\eea
Let us consider a special case, if $j_1 = j_2 =j, $ and $J$ is even, then,
second term in Eq. (\ref{direct_last}) vanishes, and the direct term becomes,

\be
D = C_0 \left( \frac{\hat{j}^4 }{2}\right)
\left (
\begin{matrix}
j&j&J\\
- \frac{1}{2}& \frac{1}{2}&0\\
\end{matrix}
\right )^2
\ee
We shall see below, for this special case, odd values of $J$ are not
allowed  for the identical nucleons. Since, the contribution from the
exchange term in this case  is  equal and opposite to that for the
direct term.

\subsubsection{Exchange term}
By substituting Eq.(\ref{del_2}) in Eq.(\ref{mat_1}) and separating
radial and angular parts of matrix elements the exchange term becomes,

\bea
E &=& (-1)^{j'_1+j'_2 +J+T} 4\pi C_0 \times \non
&& \sum_{K}
\bra{\alpha_1 j_1 \alpha_2 j_2;JM}   Y^K (1) \cdot  Y^K(2)\ket{\alpha'_2
j'_2 \alpha'_1 j'_1;JM}   \label{exchange}
\eea
It must be noticed that the radial part of the matrix element $C_0$
is same as that for the direct term  Eq.(\ref{c0}), because, the
$\delta$-interaction acts only when $\mathbf{r}_1 = \mathbf{r}_2$.
The
angular part can further be decomposed into the products of matrix
elements for $Y^K(1)$ and $Y^K(2)$ using Eq. (\ref{yk1_yk2}). The
Eq. (\ref{exchange}) simplifies to,

\bea
E&=& (-1)^{j'_1+j'_2 +J+T} 4\pi C_0 \sum_{K} (-1)^{j_2+J+j'_2}
\left \{ \begin{matrix}
j_1&j_2&J\\
j'_1&j'_2&K\\
\end{matrix} \right \} \non
&& \bra{\alpha_1 j_1} | Y^K(1) | \ket{\alpha'_2 j'_2} \bra {\alpha_2 j_2} | Y^K(2)
| \ket{ \alpha'_1 j'_1}  \non
\bra{\alpha_1; j_1}\mid Y^K\mid\ket{\alpha'_2;j'_2}
& &\bra{\alpha_2; j_2}\mid Y^K\mid\ket{\alpha'_1;j'_1} 
= (-1)^{j_1+j_2 -1}
\left( \frac{\hat{j'_1}\hat{j_1}\hat{j'_2}\hat{j_2}
\hat{K}^2}{4\pi}\right)\non
&&\quad \quad \quad \quad \quad \quad \frac{1}{4}(1+(-1)^{l'_2+l_1+K}) (1+(-1)^{l'_1+l_2+K})\non
&& \quad \quad \quad \quad\quad \quad\times   \left (
\begin{matrix}
 j_1& K&j'_2\\
-\frac{1}{2}& 0&\frac{1}{2}
\end{matrix}
\right )   \left (
\begin{matrix}
 j_2& K&j'_1\\
-\frac{1}{2}& 0&\frac{1}{2}
\end{matrix}
\right )
\nonu
\eea
For $ \alpha'_1 j'_1 =  \alpha_1 j_1$ and $ \alpha'_2 j'_2 =  \alpha_2 j_2$,

\bea
\bra{\alpha_1; j_1}\mid Y^K\mid\ket{\alpha_2;j_2}
\bra{\alpha_2; j_2}\mid Y^K\mid\ket{\alpha_1;j_1}
&&= (-1)^{j_1+j_2 -1}
\left( \frac{\hat{j_1}^2\hat{j_2}^2 \hat{K}^2}{4\pi}\right) \nonumber \\
&&
 \frac{1}{2}(1+(-1)^{l_1+l_2+K} )  \nonumber \\
&& \times \left (
\begin{matrix}
 j_1& K&j_2\\
-\frac{1}{2}& 0&\frac{1}{2}
\end{matrix}
\right )   \left (
\begin{matrix}
 j_2& K&j_1\\
-\frac{1}{2}& 0&\frac{1}{2}
\end{matrix}
\right ) \label{TME40}
\eea

\bea
E&=& (-1)^T 4\pi C_0 \left( \frac{\hat{j_1}^2\hat{j_2}^2 }{8\pi}\right)
\sum_{K} \hat{K}^2 (1+(-1)^{l_1+l_2+K})
\times \left \{ \begin{matrix}
j_1&j_2&J\\
j_1&j_2&K\\
\end{matrix} \right \}
 \left (
\begin{matrix}
 j_1& K&j_2\\
-\frac{1}{2}& 0&\frac{1}{2}
\end{matrix}
\right )   \left (
\begin{matrix}
 j_2& K&j_1\\
-\frac{1}{2}& 0&\frac{1}{2}
\end{matrix}
\right )\label{TME17}\non
&=&(-1)^T C_0 \left( \frac{\hat{j_1}^2\hat{j_2}^2 }{2}\right)
\sum_{K}\left [\hat{K}^2
\times \left \{ \begin{matrix}
j_1&j_2&J\\
j_1&j_2&K\\
\end{matrix} \right \}
 \left (
\begin{matrix}
 j_1& K&j_2\\
-\frac{1}{2}& 0&\frac{1}{2}
\end{matrix}
\right )   \left (
\begin{matrix}
 j_2& K&j_1\\
-\frac{1}{2}& 0&\frac{1}{2}
\end{matrix}
\right ) \right .\non
&&+(-1)^{l_1 +l_2 +K}\hat{K}^2
\left . \times \left \{ \begin{matrix}
j_1&j_2&J\\
j_1&j_2&K\\
\end{matrix} \right \}
 \left (
\begin{matrix}
 j_1& K&j_2\\
-\frac{1}{2}& 0&\frac{1}{2}
\end{matrix}
\right )   \left (
\begin{matrix}
 j_2& K&j_1\\
-\frac{1}{2}& 0&\frac{1}{2}
\end{matrix}
\right ) \right]\label{TME2}
\eea
The sum over $K$ can be performed analytically as in the case of direct term. Substituting $J_3 = J$, $J_3' = K$, $j'_1=j_1$ and  $j'_2=j_2$ in Eq.
(\ref{TM24}), we get,

\bea
\dss_{M_3} \left (
\begin{matrix}
j_1 &j_2& J\\
m_1& m_2& M_3\\
\end{matrix}
\right )  \left (
\begin{matrix}
j_1 &j_2& J\\
m'_1& m'_2& -M_3\\
\end{matrix}
\right )
&=&\dss_{K M'_3}(-1)^{J+K+ m_1+ m'_1} \hat{K}^2
\left \{
\begin{matrix}
j_1&j_2&J\\
j_1&j_2&K\\
\end{matrix}
\right \}\non
&& \left (
\begin{matrix}
j_1 &j_2& K\\
m'_1& m_2& M'_3\\
\end{matrix}
\right )
\left (
\begin{matrix}
j_1 &j_2& K\\
m_1& m'_2& -M'_3\\
\end{matrix}
\right )\non
&=& \dss_{K M'_3}(-1)^{J+K+ m_1+ m'_1} \hat{K}^2
\left \{
\begin{matrix}
j_1&j_2&J\\
j_1&j_2&K\\
\end{matrix}
\right \}\non
&& (-1)^ {j_2 +j_2+K} \left (
\begin{matrix}
j_1 &K&j_2\\
m'_1& M'_3& m_2\\
\end{matrix}
\right )
 \left (
\begin{matrix}
j_2 & K&j_1\\
m'_2& -M'_3& m_1\\
\end{matrix}
\right )\non
&=& \dss_{K M'_3}(-1)^{j_1+ j_2+ J+ m_1+ m'_1} \hat{K}^2
\left \{
\begin{matrix}
j_1&j_2&J\\
j_1&j_2&K\\
\end{matrix}
\right \}\non
&& \left (
\begin{matrix}
j_1 &K&j_2\\
m'_1& M'_3& m_2\\
\end{matrix}
\right )
\left (
\begin{matrix}
j_2 & K&j_1\\
m'_2& -M'_3& m_1\\
\end{matrix}
\right )\label{TME200}
\eea
Substituting $m_1 = m'_1 = -1/2 $, $m_2=m'_2= +1/2 $ and the allowed values are $ M_3= M'_3=0 $
\bea
\left (
\begin{matrix}
j_1 &j_2& J\\
 - \frac{1}{2}& + \frac{1}{2}& 0\\
\end{matrix}
\right )  \left (
\begin{matrix}
j_1 &j_2& J\\
 - \frac{1}{2}& + \frac{1}{2}& 0\\
\end{matrix}
\right )
&=& \dss_{K }(-1)^{j_1 + j_2 +J -1} \hat{K}^2
\left \{
\begin{matrix}
j_1&j_2&J\\
j_1&j_2&K\\
\end{matrix}
\right \} 
\nonumber \\ 
&&
\left (
\begin{matrix}
j_1 &K&j_2\\
- \frac{1}{2}& 0& + \frac{1}{2}\\
\end{matrix}
\right ) \left (
\begin{matrix}
j_2 & K&j_1\\
+ \frac{1}{2}& 0& - \frac{1}{2}\\
\end{matrix}
\right )
\eea
\bea
\left (
\begin{matrix}
j_1 &j_2& J\\
 - \frac{1}{2}& + \frac{1}{2}& 0\\
\end{matrix}
\right )^2 
&=&  \dss_{K }(-1)^{J+K-1} \hat{K}^2
\left \{
\begin{matrix}
j_1&j_2&J\\
j_1&j_2&K\\
\end{matrix}
\right \} \left (
\begin{matrix}
j_1 &K&j_2\\
- \frac{1}{2}& 0& + \frac{1}{2}\\
\end{matrix}
\right )
\left (
\begin{matrix}
j_2 & K&j_1\\
- \frac{1}{2}& 0&+  \frac{1}{2}\\
\end{matrix}
\right )\label{TME210}
\eea
Substituting $m_1 = m_2 = +1/2 $, $m'_1=m'_2= -1/2 $, and the allowed values of
$M_3= -1, M'_3=0$ in Eq.(\ref{TME200})
becomes,
\bea
\left (
\begin{matrix}
j_1 &j_2& J\\
  \frac{1}{2}& + \frac{1}{2}& -1\\
\end{matrix}
\right )  \left (
\begin{matrix}
j_1 &j_2& J\\
 - \frac{1}{2}& - \frac{1}{2}&1\\
\end{matrix}
\right )
&=& \dss_{K }(-1)^{j_1 +j_2 +J} \hat{K}^2
\left \{
\begin{matrix}
j_1&j_2&J\\
j_1&j_2&K\\
\end{matrix}
\right \} \left (
\begin{matrix}
j_1 &K&j_2\\
- \frac{1}{2}& 0& + \frac{1}{2}\\
\end{matrix}
\right )
\left (
\begin{matrix}
j_2 & K&j_1\\
-\frac{1}{2}& 0& + \frac{1}{2}\\
\end{matrix}
\right )\label{TME22}\non
\left (
\begin{matrix}
j_1 &j_2& J\\
 \frac{1}{2}& + \frac{1}{2}& -1\\
\end{matrix}
\right )^2
&=& \dss_{K }\hat{K}^2
\left \{
\begin{matrix}
j_1&j_2&J\\
j_1&j_2&K\\
\end{matrix}
\right \}\left (
\begin{matrix}
j_1 &K&j_2\\
- \frac{1}{2}& 0& + \frac{1}{2}\\
\end{matrix}
\right )
\left (
\begin{matrix}
j_2 & K&j_1\\
-\frac{1}{2}& 0& + \frac{1}{2}\\
\end{matrix}
\right )\label{TME230}
\eea
Substituting  Eqs. (\ref{TME210}) and (\ref{TME230}) in Eq. (\ref{TME2})
\bea
E&=&(-1)^T C_0 \left( \frac{\hat{j_1}^2\hat{j_2}^2 }{2}\right)
\left [     \left (
\begin{matrix}
j_1 &j_2& J\\
 \frac{1}{2}& + \frac{1}{2}& -1\\
\end{matrix}
\right )^2  + (-1)^{l_1+l_2+J-1}  \left (
\begin{matrix}
j_1 &j_2& J\\
- \frac{1}{2}& + \frac{1}{2}& 0\\
\end{matrix}
\right )^2  \right ]\non
&=& (-1)^T C_0 \left( \frac{\hat{j_1}^2\hat{j_2}^2 }{2}\right)
\left [ \left (
\begin{matrix}
j_1 &j_2& J\\
 \frac{1}{2}& + \frac{1}{2}& -1\\
\end{matrix}
\right )^2  - (-1)^{l_1+l_2+J}  \left (
\begin{matrix}
j_1 &j_2& J\\
- \frac{1}{2}& + \frac{1}{2}& 0\\
\end{matrix}
\right )^2  \right ]
\eea
Using the recursion relation for 3j-symbol, Eq. (\ref{rec1}),
\bea
E&=& (-1)^T
C_0 \left( \frac{\hat{j_1}^2\hat{j_2}^2 }{2}\right) 
 \left (
\begin{matrix}
j_1&j_2&J\\
-\frac{1}{2}&  \frac{1}{2}&0\\
\end{matrix}
\right )^2
\left[
 \frac{[  \hat{j_1}^2 + \hat{j_2}^2 (-1)^{j_1+ j_2+J}
]^2}{4J(J+1)}
- (-1)^{l_1+l_2+J}  
\right] \non \label{exch_last}
\eea

\subsubsection{Total matrix element}
Total matrix element is the sum of the direct and
exchange terms  with appropriate normalization factor. Using
Eqs. (\ref{mat_1}),(\ref{direct_last}), and (\ref{exch_last}), the total
matrix element can be written as,

\bea
\bra{\alpha_1 j_1 \alpha_2 j_2;JM} \delta(\vec{r})  \ket{\alpha_1 j_1
\alpha_2 j_2;JM} _{\bf{as}}
&=&
N_{12} [D+E] \non
&=& N_{12}  C_0 \left( \frac{\hat{j_1}^2\hat{j_2}^2 }{2}\right)
\left (
\begin{matrix}
j_1&j_2&J\\
- \frac{1}{2}&  \frac{1}{2}&0\\
\end{matrix}
\right )^2\non 
&& \times 
 \left [1- (-1)^{l_1 +l_2 +J +T} + (1 + (-1)^T) \frac{[\hat{j_1}^2 +
 \hat{j_2}^2 (-1)^{j_1+ j_2+J} ]^2}{4J(J+1)}
 \right ] \non \label{total}
\eea
For the special case, ${\alpha_i j_i = \alpha j}$ and $N_{12}
=\frac{1}{2}$, the Eq.(\ref{total}) becomes,
\bea
\bra{(\alpha j)^2; JM}
\delta(\vec{r})\ket{(\alpha j)^2; JM}_{as} 
&=& C_0 \left( \frac{\hat{j}^4 }{4}\right)
 \times \left (
\begin{matrix}
j&j&J\\
\frac{1}{2}&   -\frac{1}{2}&0\\
\end{matrix}
\right )^2  \left[ 1-(-1)^{J+T} +(1+ (-1)^T ) \frac{\hat{j}^4 
(1- (-1)^J)^2}{4J(J+1)}
 \right ]. \non
\eea
For the identical nucleons, $T=1$, the above matrix element is non-zero only for even values of $J$ and is  given by,
\bea
\bra{(\alpha j)^2; JM}
\delta(\vec{r})\ket{(\alpha j)^2; JM}_{as}
&=& C_0 \left( \frac{\hat{j}^4 }{2}\right)
 \times \left (
\begin{matrix}
j&j&J\\
\frac{1}{2}&   -\frac{1}{2}&0\\
\end{matrix}
\right )^2 \non
&=& C_0 \left( \frac{(2j+1)^2 }{2}\right)
 \times \left (
\begin{matrix}
j&j&J\\
\frac{1}{2}&   -\frac{1}{2}&0\\
\end{matrix}
\right )^2\nonu
\eea
Substituting the value of 3j-symbol for $J=0$ and $2$,
\be
\bra{(\alpha j)^2; JM}
\delta(\vec{r})\ket{(\alpha j)^2; JM}_{as}
= \frac{C_0}{8\pi}\left\{
\begin{array}{c}
(2j+1)\qquad\qquad \qquad \qquad  \text{for J=0}\\
(2j+1)\left (\frac{(2j-1)(2j+3)}{16j(j+1)}\right ) \qquad \text{for J=2}
\end{array}
\right . \nonu
\ee
The $\ket{(\alpha j)^2; 00}$ is referred as paired state. The matrix
element for the $\delta$-interaction between the paired states is much
larger than those between the unpaired sates i.e.,
\bea
\bra{(\alpha j)^2;00}
\delta(\vec{r})\ket{(\alpha j)^2; 00} &>>&
\bra{(\alpha j)^2;2M}
\delta(\vec{r})\ket{(\alpha j)^2; 2M}\non
\eea
The pairing Hamiltonian, as formally defined later in Section 5,  mimics the  $\delta$-interaction  acting on
the paired state so,
\be 
\bra{j^2; JM} \hat{H}_{pair} \ket{j^2;JM} \propto (2j+1)\delta_{J0}
\ee

\section{Creation and annihilation of angular momentum coupled state}

The creation and annihilation operators for coupled angular momentum
states can be conveniently expressed in terms of these operators for
single nucleon. The algebra involving these operators become simple  due
to the application of Wick's theorem as described below very briefly.
The creation operator '$a^\dagger$' and annihilation operator '$a$'
for single-nucleon can be defined as,

\bea
a^\dagger_\alpha\ket{-} =\ket{\alpha},\\
a_\beta\ket{\alpha}=\delta_{\alpha \beta}\ket{-},\\
a\ket{-}=0,
\eea
where $\ket{-}$ is the vacuum state. The anti-commutation relations for the creation and annihilation operators for single nucleon '$a^\dagger$'
and '$a$' are given by,
\bea
\left \{a^\dagger_{\alpha},a^\dagger_{\beta}\right\}=
\left \{a_{\alpha},a_{\beta}\right\}&=&0\non
\left \{a^\dagger_{\alpha},a_{\beta}\right\}&=&\delta_{\alpha \beta}
\eea
The normal order product  is defined as,
\bea
:a_\beta a^\dagger_\gamma: &=& -a^\dagger_\gamma a_\beta\\
:a_\beta a^\dagger_\gamma a^\dagger_\delta: &=& a^\dagger_\gamma
a^\dagger_\delta a_\beta\\
:a_\alpha  a_\beta a^\dagger_\gamma a^\dagger_\delta: &=& a^\dagger_\gamma
a^\dagger_\delta a_\alpha  a_\beta\\
\bra{-}:a_\alpha  a_\beta a^\dagger_\gamma a^\dagger_\delta: \ket{-}&=&
\bra{-}a^\dagger_\gamma a^\dagger_\delta a_\alpha  a_\beta\ket{-} =0 
\eea
The normal ordered product essentially arranges all the creation operators
left to the annihilation operators as should be  clear from the above
equations. The over all sign is negative  only if number of permutations
required to shift all the creation operators to the left is odd.

The contractions  of two operators are given by,
\bea
\ob{a_\beta a^\dagger_\gamma}&=&  \bra{-}a_\beta a^\dagger_\gamma\ket{-} =
\delta_{\beta \gamma}\non
\ob{a^\dagger_\gamma a_\beta}&=&  \bra{-} a^\dagger_\gamma a_\beta\ket{-}   =0\non
\ob{a_\alpha a_\beta} &=& \ob{a^\dagger_\gamma a^\dg_\delta}=0\non
\eea
A product of set of  creation and annihilation operators can be simplified
in terms of normal order products and contractions using Wick's
theorem. For example, the product of a set of four creation and annihilation
operators can be expressed using  the Wick's theorem  as,
\bea
a_\alpha a_\beta a^\dg_\gamma a^\dg_\delta &=& : a_\alpha a_\beta
a^\dg_\gamma a^\dg_\delta: +\sum_{singles}\ob{aa^\dg}: aa^\dg:
+\sum_{doubles}\ob{aa^\dg}\ob{aa^\dg} \label{wick}
\eea
The sums in the above equation run over all the possible single and
double contractions. The Eq.(\ref{wick}) can easily be extended to a
product of any number of creation and annihilation operators.

\subsection{Coupled states}
The coupled state of two nucleons can be expressed in terms of
product of the single-nucleon states as  \cite{Talmi93,Pal82,Heyde90,Casten00},

\bea
\ket{j_1 j_2;JM}&=& \sum_{m_1,m_2}\braket{j_1 m_1 j_2
m_2}{j_1 j_2 JM}\ket{j_1m_1 j_2m_2}
\eea
where, $j_i$  and $m_i$ are the angular  momentum and its $z$-component
for the single-nucleon, respectively. The  $ \braket{j_1 m_1 j_2
m_2}{j_1 j_2 JM} $ is the Clebsch-Gordan coefficient. The
anti-symmetrization of the  above equation yields,
\bea
\ket{j_1 j_2;JM}_{as} 
& =&{\mathcal{N}} \sum_{m_1,m_2}\braket{j_1 m_1 j_2
m_2}{ j_1 j_2 JM}( \ket{j_1m_1j_2m_2} -\ket { j_2m_2j_1m_1 })\non
&=&  {\mathcal{N}} \sum_{m_1,m_2}\braket{j_1 m_1 j_2
m_2}{j_1 j_2 JM} \ket{  j_1m_1j_2m_2 }_{as}\label{PO_4}
\eea
The operators $A^\dagger(j_1,j_2;JM)$ and $A(j_1,j_2;JM)$ which create and annihilate two nucleons in a coupled state can be expressed analogous to Eq. (\ref{PO_4}) as,
\bea 
{A^\dg} ( j_1 j_2;JM) &=& {\mathcal{N}} 
\sum_{m_1,m_2}\braket{j_1 m_1 j_2
m_2}{j_1 j_2 JM} a^\dg_{j_1m_1} a^\dg_{j_2m_2} \label{A_dag}\\
A ( j_1 j_2;JM)  &=& {\mathcal{N}}  \sum_{m_1,m_2}\braket{j_1 m_1
j_2m_2}{j_1 j_2 JM}   a_{j_2m_2} a_{j_1m_1} \label{A}
\eea
\subsubsection{Normalization constant}
The normalization constant can easily be calculated by applying Wick's theorem  such that,
\bea
\bra{}A( j_1 j_2;JM)  A^\dagger( j_1 j_2;JM)  \ket{-}=1.
\eea
Using Eqs. (\ref{A_dag}) and (\ref{A}),
\bea
\bra{-}A ( j_1 j_2;JM)  A^\dg ( j_1 j_2;JM)\ket{-} &=&
\mathcal{N}^2 \sum_{m_1m_2}\sum_{m_1'm_2'} \braket{j_1 m_1 j_2 m_2}{ j_1
j_2 JM}\non
&& \braket{j_1 m'_1 j_2
m'_2}{JM}\bra{-} a_{j_2m'_2} a_{j_1m'_1} a^\dg_{j_1m_1}
a^\dg_{j_2m_2}\ket{-} \non
\label{PO_3}
\eea
The R.H.S. of the Eq.(\ref{PO_3}) can be simplified by applying the Wick's
theorem Eq.(\ref{wick}). Also, the vacuum expectation value for a normal ordered product vanishes. Only the fully contracted terms
will contribute, we have
\bea
\bra{-} a_{j_2m'_2} a_{j_1m'_1} a^\dg_{j_1m_1} a^\dg_{j_2m_2}\ket{-} &=&
\bra{-}\left \{ \ob{a_{j_2m'_2} a^\dg_{j_2m_2}} \ob{a_{j_1m'_1}
a^\dg_{j_1m_1} }- \ob {a_{j_2m'_2}a^\dg_{j_1m_1}} \ob{a_{j_1m'_1}
a^\dg_{j_2m_2}} \right \} \ket{-}\non
\bra{-} a_{j_2m'_2} a_{j_1m'_1} a^\dg_{j_1m_1} a^\dg_{j_2m_2}\ket{-} &=&
\delta_{m_1 m_1'}\delta_{m_2 m_2'} - \delta_{j_1j_2}\delta_{m_1 m_2'}
\delta_{m_2 m'_1}
\label{PO_5}.
\eea
Substituting  Eq. (\ref{PO_5}) in Eq. (\ref{PO_3}),
\bea
\bra{-}A ( j_1 j_2;JM)  A^\dg ( j_1 j_2;JM)\ket{-} &=&
\mathcal{N}^2 \sum_{m_1m_2}\sum_{m_1'm_2'} \braket{j_1 m_1 j_2 m_2}{j_1 j_2
JM}
\braket{j_1 m'_1 j_2 m'_2}{j_1 j_2 JM} \non
&& ( \delta_{m_1 m'_1}\delta_{m_2 m'_2} - \delta_{j_1j_2}\delta_{m_1 m'_2}
\delta_{m_2 m'_1}) \non
&=&\mathcal{N}^2  \sum_{m_1m_2} \left(\braket{j_1 m_1 j_2 m_2}{j_1 j_2 JM}
 \braket{j_1 m_1 j_2 m_2}{j_1 j_2 JM} \right. \non
&&\left . -\delta_{j_1 j_2} \braket{j_1 m_1 j_2 m_2}{j_1 j_2 JM}
 \braket{j_1 m_2 j_2 m_1}{j_1 j_2 JM}   \right) \non
&=&\mathcal{N}^2\sum_{m_1m_2} \left(\braket{j_1 m_1 j_2 m_2}{j_1 j_2 JM}^2 \right.
\non
&&\left . -\delta_{j_1 j_2}(-1)^{j_1+j_2-J} \braket{j_1 m_1 j_2 m_2}{j_1
j_2 JM}
 \braket{j_2 m_1 j_1 m_2}{j_1 j_2 JM}   \right)\non
&=&\mathcal{N}^2  \sum_{m_1m_2} \left(\braket{j_1 m_1 j_2 m_2}{j_1 j_2 JM}^2   + \delta_{j_1 j_2} (-1)^{-J} \braket{j_1 m_1 j_2 m_2}{j_1 j_2 JM}^2 \right)  \label{phase} \\
\bra{-}A ( j_1 j_2;JM)  A^\dg ( j_1 j_2;JM)\ket{-}
&=&\mathcal{N}^2 (1+\delta_{j_1 j_2} (-1)^{-J})\sum_{m_1m_2}
 \braket{j_1 m_1 j_2 m_2}{j_1 j_2 JM}^2\non
\bra{-}A ( j_1 j_2;JM)  A^\dg ( j_1 j_2;JM)\ket{-}
&=&\mathcal{N}^2 (1+\delta_{j_1 j_2} (-1)^{-J})\non
\mathcal{N} &=& \frac{1}{\sqrt{1+\delta_{j_1 j_2}}}\label{norm} 
\eea
In Eq.(\ref{phase}) we have  substituted  $j_1=j_2$ in the phase factor
due to the  presence of $\delta_{j_1j_2}$.  It may be easily verified
from Eq. (\ref{PO_4})  that if $j_1 = j_2$, only the even value of $J$
are allowed.

\section{Pairing Hamiltonian}
Two nucleons are said to be in the paired state when, $j_1=j_2=j,
J=M=0$.  The creation and annihilation operators for  the paired state
can be obtained using Eqs. (\ref{A_dag}) and (\ref{A}) together with
(\ref{norm}),i.e.,
\bea
{A^\dg} ( jj;00) &=&  \frac{1}{\sqrt{2}}
\sum_{m_1,m_2}\braket{j m_1
jm_2}{jj00}  a^\dg_{jm_1} a^\dg_{jm_2} \non
&=&  \frac{1}{\sqrt{2}} \sum_m \frac{(-1)^{j-m}}{\sqrt{2j+1}}
  a^\dg_{jm}a^\dg_{j,-m}\label{PO_7}\\
 A(jj:00)  &=&  \frac{1}{\sqrt{2}} \sum_m \frac{(-1)^{j-m}}{\sqrt{2j+1}}
  a_{j,-m}a_{jm}\label{PO_8}
\eea
In the above Eqs. (\ref{PO_7}) and (\ref{PO_8}) the operators $A^\dagger(jj;00)$
creates a pair while $A(jj;00)$ annihilates it. To facilitate the
discussion of pairing Hamiltonian, the operators $A^\dagger$ and $A$
can further be simplified as,

\bea
{A^\dg} ( jj;00)  &=& \sqrt{\frac{1}{\Omega}} \sum_{m>0} (-1)^{j-m}
a^\dg_{jm}a^\dg_{j,-m} \label{PO_9}  \\ 
A(jj:00) &=&\sqrt{\frac{1}{\Omega}}  \sum_{m>0} (-1)^{j-m}
a_{j,-m}a_{jm} \label{PO_10}
\eea
where, $\Omega = \frac{1}{2}(2j+1)$ is the pair degeneracy or the maximum
number of pairs in a given $j$-shell.

The pairing Hamiltonian in a single $j$-shell can be defined as,
\bea
\hat{H}_{pair}&=&-G S_j^+ S_j^-\label{hpair}
\eea
where, $G$ is the pairing strength. The operators $S_j^+ S_j^-$ are the
quasi-spin operators as they obey the commutation rules analogous to those for the angular
momentum operators (see Appendix  \ref{A1}).  These operators  can be expressed in terms of the pair creation
and annihilation operators as,
\bea
S^+_j&=& \sqrt{\Omega}A^\dagger  = \sum_{m>0} (-1)^{j-m}
 \adag_{jm}\adag_{j,-m} \label{s+}\\
S^-_j &=& \sqrt{\Omega}A = \sum_{m>0} (-1)^{j-m} a_{j,-m}a_{jm} \label{s-}
\eea
Using Eqs. (\ref{s+}) and (\ref{s-}), the pairing Hamiltonian can be
written as,

\bea
\hat{H}_{pair}&=&-G \dss_{m,m' > 0}(-1)^{2j-m-m'}
 \adag_{jm}\adag_{j,-m}a_{j,-m'}a_{jm'}\nonu
\eea
It can be easily verified that the matrix element or the pairing
Hamiltonian is non-zero only between the paired states, i.e.,  
\bea
\bra{j^2; JM} \hat{H}_{pair} \ket{j^2;JM} = \frac{-G}{2} (2j+1)\delta_{J0}
\delta_{M0}. 
\eea

\subsection{Quasi-spin and seniority}
The eigen values of the pairing Hamiltonian can be obtained once
the commutation relation between $S^+_j$ and $S^-_j$ is known.
In what follows, we drop the index $j$ from the operators $S^+_j$ and
$S^-_j$. This commutation relation can be easily evaluated by applying
Wick's theorem, as briefly described in Appendix \ref{A1}, one gets,
\bea
\left[ S^+, S^-\right ]_- = \hat{N} - \Omega = 2S^0
\eea
where $S^0$ is a operator analogous to z-component of quasi spin in a way that $S^+, S^-$ and $S^0$ follow the SU(2) Lie algebra (Appendix \ref{A1}). Also, the particle number operator ($\hat{N}$),
\be
\hat{N} =\dss_{m > 0} [a^\dg_m  a_m+ a^\dg_{-m} a_{-m}] .
\ee
It can be further shown that,
\bea
[S^+,S^0]_- = -S^+\\
\left[S^-,S^0\right ]_-=S^-
\eea
The  commutations of $S^+, S^-$ and $S^0$ suggest that these
operators follow the angular momentum algebra and can be called as
quasi-spin operators.
Therefore,

\bea
S^+S^- &=&{\bf{S}}^2-(S^0)^2+S^0\non
&=& S(S+1)- S^0(S^0-1)\non
&=& S(S+1)-\left( \frac{\hat{N}-\Omega }{2} \right )\left(
\frac{\hat{N}-\Omega}{2} -1 \right)\non
&=& S(S+1)-\left( \frac{\Omega-\hat{N}}{2} \right )\left(
\frac{\Omega- \hat{N}}{2} +1 \right)
\label{PH_19}
\eea
where $S(S+1)$ is  the eigen-value for the operator ${\bf{S}}^2$ with
$S$  being an integer or half-integer analogous to the case of angular
momentum.  In principle, the eigen-state of the pairing Hamiltonian
can be characterized by the  quasi-spin $S$ and its $z$-component
$S^0$. However, it is convenient to express them in terms of the
total number of nucleons and the seniority quantum number, i.e., number
of the unpaired nucleons.

Let us consider a $n$-nucleon state $\ket{j^n,\nu;JM}$ with $\nu$ number of unpaired nucleons or the seniority quantum number.
Thus,  if we consider a state with all the nucleons unpaired,
i.e., $n=\nu$, $\hat{N}\ket{j^\nu,\nu;JM}=\nu\ket{j^\nu,\nu;JM}$  and
$S^-_j\ket{j^\nu,\nu;JM}=0$, since, there is no pair to annihilate.
By operating L.H.S. and R.H.S. of Eq.(\ref{PH_19}) on  $\ket{j^\nu,\nu;JM}$,
\bea
 S(S+1)-\left( \frac{\Omega-\hat{N}}{2} \right )\left(
\frac{\Omega- \hat{N}}{2} +1 \right)\ket{j^{\nu}\nu JM}
 &=& S^+ S^-\ket{j^{\nu}\nu JM}\non
S(S+1)-\frac{ \Omega -\nu}{2} \left(\frac{ \Omega -\nu}{2} +1 \right)&=&0
\non
S &=& \frac{ \Omega -\nu}{2}\label{PH_20}\non
\eea
{Substituting Eq. (\ref{PH_20})  in Eq. (\ref{PH_19}), }
\bea
S^+ S^-&=& \frac{ \Omega -\nu}{2}\left(\frac{ \Omega -\nu}{2}+1\right)
- \left( \frac{\Omega-\hat{N}}{2} \right )\left(
\frac{\Omega- \hat{N}}{2} +1 \right)\non
 S^+ S^-&=& \frac{ 1}{4} (\hat{N} -\nu )( 2 \Omega  -\hat{N}-\nu +2)
\label{PH_21} \eea

\subsection{Eigen values of $ \hat{H}_{pair}$ }
The pairing Hamiltonian using Eqs. (\ref{hpair}) and (\ref{PH_21}) becomes,\\
\bea
\hat{H}_{pair} &=& -G  S^+ S^- =   \frac{ -G}{4} (\hat{N} -\nu )
( 2 \Omega  -\hat{N}-\nu +2) \nonu 
\eea
\bea
\hat{H}_{pair}  \ket{j^n,\nu;JM} &=&  \frac{-G }{4} (\hat{N} -\nu )
( 2 \Omega  -\hat{N}-\nu +2)\ket{j^n,\nu;JM}\non
&=&  \frac{ -G}{4} ({n} -\nu )( 2 \Omega  -{n}-\nu +2)\ket{j^n,\nu;JM}\non
 &=&E(n,\nu) \ket{j^n,\nu;JM}\nonu
 \eea
The pairing eigen value $ E(n,\nu)$ can be obtained as, 
 \be
E(n,\nu) =  \frac{ -G}{4} ({n} -\nu )( 2 \Omega  -{n}-\nu +2)\nonumber
\ee
For fully paired state, $\nu=0$, in case of even-even nuclei
\be
 E(n,0) =  \frac{-G}{4} {n} (2 \Omega  -{n}+2)\nonu
\ee
The pairing energy  is given as,
\be
 E(n,\nu) - E(n,0)  =
 \frac{G}{4} {\nu} (2 \Omega  -{\nu}+2)  
\label{epair}\ee
It may be noted that the pairing energy is independent of the number of
nucleons in the $j$-shell.

\subsection{Simple applications of Pairing}

\subsubsection{Even-Even nuclei}
\begin{figure}
\includegraphics[width=1\columnwidth,angle=0,clip=true]{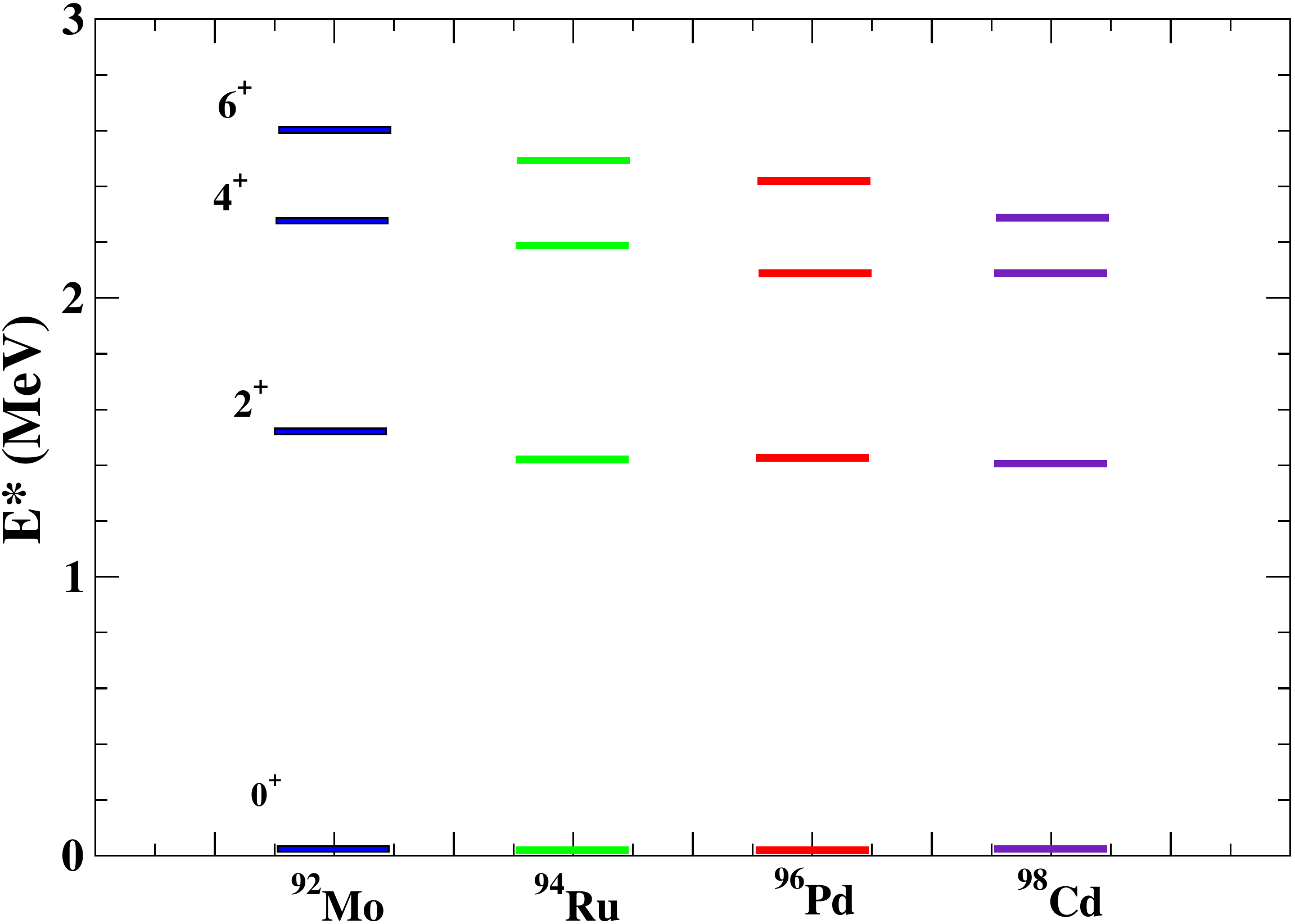}
\caption{\label{fig3} Energy spectrum of even-even nuclei from mass-90 region.} 
\end{figure}

One can understand  the energy separation of $0^+$ and
$2^+$ in even-even nuclei in terms of pairing. For instance, in Fig.
\ref{fig3} we display the observed spectra for a few nuclei around
mass number $A =  90$ with the neutron number $N=50$. It may be noted
that the energy separations  between the $0^+$ and $2^+$ for these
nuclei are almost $1.4-1.5$ MeV. This is the energy usually required
to break one nucleon pair in this mass region. Ground-state
for the even-even nuclei corresponds to fully paired state i.e., $J =
0^+ $, $ \nu =0 $. First excited-state  corresponds to breaking of one
proton pair yielding  $J = 2^+ $,  $\nu =2 $.  Therefore, within the
pairing model, the energy separation between the ground and the first
excited states can be approximated using Eq.(\ref{epair}) as,

\bea
E(2^+)-E(0^+) &\approx & E(n,2)-E(n,0)\non
E(n,2)-E(n,0) &=&G \Omega.  
\eea
For $A=90$ mass region, valence nucleons are in orbit $ g_{9/2}$, i.e.,  
$ \Omega =5$, typical value of  $G= \frac{25}{A}$ yields,
\be
E(n,2)-E(n,0) \sim 1.4 \text{ MeV}
\ee

\subsubsection{Odd-A nuclei}
Consider $^{43}$Ca with three particles in $j=7/2$ orbital outside the
closed shell. How do these three angular momenta $j$ couple to give final
total $J$ values? The easiest is to use m-scheme. If we use m-scheme for
three particles in $j=7/2$ then the allowed $J$ values are $15/2, 11/2,
9/2, 7/2, 5/2, 3/2$. For the case of $J=7/2$ two of the particles must
have their angular momenta coupled to $J=0$, giving a total $J=7/2$
from ${(7/2)}^3$ configuration. For the $J=15/2,11/2,9/2,5/2$ and
$3/2$, there are no pairs of particles coupled to $J=0$. Since a $J=0$
pair is the lowest configuration for two particles in the same orbital,
$J=7/2$ must lie the lowest for three particles in the same orbital. From
$^{41}$Ca to $^{47}$Ca, odd-A isotopes, ${7/2}^-$ state lies as the ground
states. These cases are still simple to understand, since as the number
of valence nucleons grows, the number of ways of mixing basis states to
generate a given $J$ increases rapidly. So, the large-scale shell model
calculations are trending in recent times. However, pairing provides a simple physics understanding of such nuclear features.

\section{Summary}

We have presented detailed derivation for the matrix elements of
the $\delta$-interaction evaluated between anti-symmetrized coupled
states. The multipole expansion method used for this purpose enables
one to decompose the matrix elements into their respective radial and
angular parts.   The angular part of the matrix elements are evaluated
analytically and the similar procedure can be followed for different
interactions.  The standard pairing Hamiltonian is then considered
whose matrix elements between the paired states are similar to those for
the $\delta$-interaction. The solution for the pairing Hamiltonian is
obtained within the quasi-spin formalism. Finally, simple application
of the pairing to explain the energy separation between the lowest  $0^+$
and $2^+$ states in the even-even nuclei and the ground state spin for
the odd-A nuclei are discussed. Some important relations involving  matrix
elements for the spherical harmonics and the commutation relations for the
quasi-spin operators are presented in the Appendices in order to make the
derivation self-contained.  The quasi-spin algebra is performed directly
in terms of the creation and annihilation operators for single-nucleon
together with the application of the Wick's theorem.  These techniques
are quite useful in dealing with the mean-field equations for the  more
general cases such as the Hartree-Fock, Bardeen–Cooper–Schrieffer
and Hartree-Fock Bogoliubov theories.

\begin{appendices}
\section {Commutation of  $S^+,S^-$ and $S^0$  operators} \label{A1}
The commutation of  $S^+$ and $S^-$ can be expressed using Eqs. (\ref{s+})
and (\ref{s-}) as,

\bea
\left[ S^+,S^-\right]_-  &=&\dss_{m,m' > 0} (-1)^{2j-m-m'} [
\adag_{m}\adag_{-m},a_{-m'}a_{m'}]_-\non
 &=&\dss_{m,m' > 0} (-1)^{2j-m-m'} (
\adag_{m}\adag_{-m}a_{-m'}a_{m'} -
a_{-m'}a_{m'}\adag_{m}\adag_{-m}). \label{PH_6}
\eea
By applying  Wick's theorem (Eq. (\ref{wick})) to the 2nd term in Eq.(\ref{PH_6}), 
\bea 
a_{-m'} a_{m'} a^\dg_m  a^\dg_{-m}&=&
a^\dg_m a^\dg_{-m} a_{-m'} a_{m'} \non 
&&+\delta_{-m' m} a^\dg_{-m}  a_{m'}
- \delta_{-m',-m} a^\dg_m  a_{m'}
- \delta_{m' m}  a^\dg_{-m} a_{-m'}
+ \delta_{m', -m}a^\dg_{m}  a_{-m'} \non
&&-\delta_{-m'm}\delta_{m',-m}  + \delta_{-m',-m}
\delta_{m' m}. \label{PH_11}
\eea
The terms involving single and double Kronecker-delta  are due to the
single and double  contractions, respectively.
Substituting Eq. (\ref{PH_11}) into Eq. (\ref{PH_6}) we get,
\bea
\left[ S^+,S^-\right]_-
&= &\dss_{m,m' > 0} (-1)^{2j-m-m'} [
-\delta_{-m' m} a^\dg_{-m}  a_{m'}
 +\delta_{-m',-m} a^\dg_m  a_{m'}
+ \delta_{m' m}  a^\dg_{-m} a_{-m'}\non
&&- \delta_{m', -m}a^\dg_{m}  a_{-m'}
+\delta_{-m'm}\delta_{m',-m}  - \delta_{-m',-m}
\delta_{m' m}] 
\eea
Since sum runs over $m,m'>0$ ,
\bea
\left[ S^+,S^-\right]_-&=& \dss_{m,m' > 0} (-1)^{2j-m-m'} [
+\delta_{-m',-m} a^\dg_m  a_{m'}
+ \delta_{m' m}  a^\dg_{-m} a_{-m'}
 - \delta_{-m',-m} \delta_{m' m}]\non
&=& \dss_{m > 0} (-1)^{2j-2m} [a^\dg_m  a_m+ a^\dg_{-m} a_{-m} -1]\non
&=& \dss_{m > 0} [a^\dg_m  a_m+ a^\dg_{-m} a_{-m} -1]\non
\left[ S^+,S^-\right]_-&=& \hat{N} - \Omega \label{PH_12} \\
\left[ S^+,S^-\right]_-&=& 2S^0
\eea
where $\hat{N} = a^\dg_m  a_m+ a^\dg_{-m} a_{-m}. $  Similarly, one can show that, 
\bea
{[S^+,S^0]_- = -S^+}; \nonu
[S^-,S^0]_-=S^- \nonu
\eea
The commutations of $S^+, S^-$ and $S^0$ suggest that these operators
follow the angular momentum algebra and can be called as quasi-spin
operators.

\section { Important identities/relations}

One often encounters the  matrix elements  of the scalar product of
two spherical harmonics to be evaluated between the coupled states. Such
matrix elements are conveniently expressed in terms of the 6j-symbol
and the product of the reduced matrix elements of the  each of the
spherical harmonics.  The 6j-symbol contains the geometrical factor of the
angular momentum couplings, whereas, the physics is contained in the reduced
matrix elements. Following  Eq.(10.27) of Ref. \cite{Talmi93}, 
\begin{enumerate}
\item Matrix element of $Y^K \cdot Y^K$
\bea
\bra{\alpha_1  j_1  \alpha_2 j_2;JM}& Y^K(1) .Y^K(2) &\ket{\alpha'_1 j'_{1}
\alpha'_2  j'_{2};JM}
= (-1)^{j_2+J+j'_{1}} \left \{
\begin{matrix}
j_1 & j_2 &J \\
j'_{2} & j'_{1} & K
\end{matrix}
\right \} \non
&& (\alpha_1 j_1 || Y^K(1) || \alpha'_1 j'_{1})
(\alpha_2 j_2 ||Y^K(2)||\alpha'_2 j'_{2})\label{yk12} \\
(\alpha_1 j_1 || Y^K || \alpha'_1 j'_{1}) & \equiv & (\frac{1}{2} l_1 j_1
||Y^K|| \frac{1}{2} l'_1 j'_1 )\non
( \frac{1}{2} l_1 j_1 ||Y^K|| \frac{1}{2} l'_1 j'_1 )
&=&(-1)^{j_1 -1/2} \frac{\hat{j_1} \hat{j'_1} \hat{K}}{\sqrt{4\pi }}  \non
&&\times \frac{1}{2} [ 1+ (-1)^{l_1+l'_1+ K} ]
\left ( \begin{array}{ccc}
j_1 & K& j'_1\\
- \frac{1}{2} &0 & \frac{1}{2}  \\
 \end{array} \right )\label{yk1_yk2}
\eea
where, $\hat{x}=\sqrt{2x+1}$.
\item Recursion relation for 3j-symbols
\bea
\left (
\begin{matrix}
j_1&j_2&J\\
\frac{1}{2}&  \frac{1}{2}&-1\\
\end{matrix}
\right ) &=&
-\frac{1}{2\sqrt{J(J+1)}}[ \hat{j}^2_1 + \hat{j}^2_2 (-1)^{j_1+ j_2+J}]
 \left (
\begin{matrix}
j_1&j_2&J\\
-\frac{1}{2}&   \frac{1}{2}&0\\
\end{matrix}
\right ) \label{rec1} \non
\eea
\end{enumerate}

\section{Matrix elements of spherical harmonics}

We provide some details of the  derivation for  the matrix elements of
the spherical harmonics (see Eq.  (\ref{yk1_yk2})).  By using Wigner-Eckart's
theorem \cite{Talmi63,Talmi93,Pal82,Heyde90,Casten00},

\begin{eqnarray}
 \langle lm|Y^{LM}|l'm' \rangle &=&\int Y^{lm*} Y^{LM} Y^{l'm'} d\tau \nonumber \\ 
&=& {(-1)}^{l-m} \begin{pmatrix} l & L & l' \\ -m & M & m' \end{pmatrix}  \langle l|| Y^L || l' \rangle 
\end{eqnarray}
Also, the scalar product of spherical harmonics,
\begin{eqnarray}
Y^l.Y^{l'}=\sum_{m} {(-1)}^m Y^{lm}(\theta, \phi) Y^{l',-m'} (\theta', \phi') 
\end{eqnarray}
is invariant with respect to the rotation of axes. It follows that
the scalar product must be a function of angle $\Theta$ between the
directions $(\theta, \phi)$ and $(\theta', \phi')$. This angle $\Theta$
is the only quantity independent of the choice of axes. Choosing axes
so that the direction $(\theta', \phi')$ becomes the new $z-$ axis,

\begin{eqnarray}
Y^{lm}(\theta, \phi) \rightarrow Y^{lm}(0,0) = \delta_{m,0} \\
Y^{l0}(\theta', \phi') \rightarrow \hat{l} \sqrt{\frac{1}{4\pi}} P_l(cos \theta) = \hat{l} \sqrt{\frac{1}{4\pi}}P_l(cos \Theta) \\
Y^l.Y^{l'}= \hat{l} \sqrt{\frac{1}{4\pi}} \hat{l'}\sqrt{\frac{1}{4\pi}} P_l(cos\Theta)
\end{eqnarray} 
where $\hat{l}=\sqrt{2l+1}$ and $\hat{l'}=\sqrt{2l'+1}$. Also, if $Y^{lm}(\theta, \phi)$ and $Y^{l'm'}(\theta', \phi')$ are
spherical harmonics of the same angles $(\theta, \phi)$ then $\sum_{mm'}
\langle LM|lml'm' \rangle \hat{l} \sqrt{\frac{1}{4\pi}} \hat{l'} \sqrt{\frac{1}{4\pi}}
Y^{lm}(\theta, \phi)Y^{l'm'}(\theta, \phi)$ is a tensor of rank
$L$. That is,

\begin{eqnarray}
&\sum_{mm'} \langle LM|lml'm' \rangle & \hat{l} \sqrt{\frac{1}{4\pi}} \hat{l'} \sqrt{\frac{1}{4\pi}}Y^{lm}(\theta, \phi)Y^{l'm'}(\theta, \phi) = A^L \hat{L} \sqrt{\frac{1}{4\pi}} Y^{LM}(\theta, \phi) \nonumber 
\\
&& = \langle L0|l0l'0 \rangle \hat{L} \sqrt{\frac{1}{4\pi}} Y^{LM}(\theta, \phi)
\end{eqnarray}
where $\hat{L}=\sqrt{2L+1}$ and the value of $A^L$ can be found by putting $\theta=0$, $\phi=0$.
\begin{eqnarray}
\sum_{m,m'} \langle L M | l m l' m' \rangle \hat{l} \sqrt{\frac{1}{4\pi}} \hat{l'} \sqrt{\frac{1}{4\pi}} Y^{lm}(0,0) Y^{l'm'}(0,0) = A^L  \hat{L} \sqrt{\frac{1}{4\pi}} Y^{LM}(0,0) \nonumber \\
\sum_{m,m'} \langle L M | l m l' m' \rangle  \delta_{m,0} \delta_{m',0} = A^L \delta_{M,0} \nonumber \\
 \langle L 0 | l 0 l' 0 \rangle  = A^L \nonumber 
\end{eqnarray}
where $|l-l'| \le L \le |l+l'|$, and $M=m+m'$.  
So,
\begin{eqnarray}
&Y^{LM}(\theta, \phi) &= \sum_{m,m'} \frac{\hat{l}\hat{l'}}{\hat{L}}
\sqrt{\frac{1}{4\pi}} \langle LM|lml'm' \rangle  \langle L0|l0l'0\rangle Y^{lm}(\theta, \phi) Y^{l'm'}(\theta,\phi) \nonumber \\
 &Y^{lm}(\theta, \phi)& Y^{l'm'}(\theta,\phi)= \sum_{M} \langle lm l'm' |LM \rangle \langle l0 l'0 |L0 \rangle \frac{\hat{l}\hat{l'}}{\hat{L}} \sqrt{\frac{1}{4\pi}} Y^{LM}(\theta,\phi) \nonumber \\
 &Y^{LM}(\theta, \phi)& Y^{l'm'}(\theta,\phi) = \sum_{m} \langle LM l'm' |lm \rangle \langle L0 l'0 |l0 \rangle \frac{\hat{L}\hat{l'}}{\hat{l}} \sqrt{\frac{1}{4\pi}} Y^{lm}(\theta,\phi) \nonumber
\end{eqnarray}
In terms of 3j-symbols,
\begin{eqnarray}
Y^{LM}(\theta, \phi) Y^{l'm'}(\theta,\phi) &=& \sum_{m} {(-1)}^m \begin{pmatrix}
l' & L & l \\ m' & M & -m
\end{pmatrix} \begin{pmatrix} l' & L & l \\ 0 & 0 & 0
\end{pmatrix}  \nonumber \\ && \hat{L} \hat{l}\hat{l'}\sqrt{\frac{1}{4\pi}} Y^{lm}(\theta,\phi) 
\end{eqnarray}
This implies that
\begin{eqnarray}
\int Y^{lm*} Y^{LM} Y^{l'm'} d \tau & = & \int Y^{lm*} {(-1)}^m \begin{pmatrix}
l' & L & l \\ m' & M & -m
\end{pmatrix} \nonumber \\ && \begin{pmatrix} l' & L & l \\ 0 & 0 & 0
\end{pmatrix} \hat{L} \hat{l}\hat{l'} \sqrt{\frac{1}{4\pi}} Y^{lm} d\tau \\
&=& {(-1)}^m \begin{pmatrix}
l' & L & l \\ m' & M & -m
\end{pmatrix} \begin{pmatrix} l' & L & l \\ 0 & 0 & 0
\end{pmatrix}  \nonumber \\ && \hat{L} \hat{l}\hat{l'} \sqrt{\frac{1}{4\pi}} 
\end{eqnarray}
The last step is due to orthogonality of spherical harmonics.
Therefore,
\begin{eqnarray}
\langle l||Y^L||l' \rangle = {(-1)}^l \hat{L} \hat{l}\hat{l'}\sqrt{\frac{1}{4\pi}} \begin{pmatrix} l' & L & l \\ 0 & 0 & 0
\end{pmatrix}
\end{eqnarray}
In $| l j \rangle$ -basis, by using Wigner-Eckart theorem,
\begin{eqnarray}
 \langle l j || Y^L || l j'\rangle &=& \langle \frac{1}{2} l j || Y^L || \frac{1}{2}l' j' \rangle \nonumber \\
&=& {(-1)}^{\frac{1}{2}+l'+j+L} \hat{j} \hat{j'} \nonumber \\ && \begin{Bmatrix} l & j & \frac{1}{2} \\ j' & l' & L 
\end{Bmatrix} \langle l|| Y^L|| l' \rangle \quad
\end{eqnarray}
where 
\begin{eqnarray}
\langle l|| Y^L|| l' \rangle= {(-1)}^l \hat{L} \hat{l} \hat{l'}\sqrt{\frac{1}{4\pi}} \begin{pmatrix} l' & L & l \\ 0 & 0 & 0
\end{pmatrix} \nonumber
\end{eqnarray}
Therefore, we need to calculate the term $ \begin{Bmatrix} l & j & \frac{1}{2} \\ j' & l' & L 
\end{Bmatrix} \begin{pmatrix} l' & L & l \\ 0 & 0 & 0
\end{pmatrix}$:
\begin{eqnarray}
\begin{Bmatrix} l & j & \frac{1}{2} \\ j' & l' & L 
\end{Bmatrix} \begin{pmatrix} l' & L & l \\ 0 & 0 & 0
\end{pmatrix} &=& \begin{Bmatrix} l' & L & l \\ j' & \frac{1}{2} & j 
\end{Bmatrix} \begin{pmatrix} l' & L & l \\ 0 & 0 & 0
\end{pmatrix} \nonumber \\
&=& \sum_{all m's} {(-1)}^{j+j'+\frac{1}{2}+m_s+m_j+m_j'} \begin{pmatrix} l' & L & l \\ 0 & 0 & 0
\end{pmatrix} \nonumber \\ &&  \begin{pmatrix} l' & \frac{1}{2} & j' \\ 0 & m_s & -m_j'
\end{pmatrix} \begin{pmatrix} j & L & j' \\ -m_j & 0 & m_j'
\end{pmatrix}
 \nonumber \\ &&  \begin{pmatrix} j & \frac{1}{2} & l \\ m_j & -m_s & 0
\end{pmatrix}\begin{pmatrix} l' & L & l \\ 0 & 0 & 0
\end{pmatrix}
\end{eqnarray} 
Since, $\sum_{m'_l m_L m_l} \begin{pmatrix} l' & L & l \\ 0 & 0 & 0
\end{pmatrix}\begin{pmatrix} l' & L & l \\ 0 & 0 & 0
\end{pmatrix}=1$, however, $ m'_l=m_L=m_l=0$. Therefore,

\begin{eqnarray}
\begin{Bmatrix} l & j & \frac{1}{2} \\ j' & l' & L 
\end{Bmatrix} \begin{pmatrix} l' & L & l \\ 0 & 0 & 0
\end{pmatrix} &=& \sum_{m_sm_jm_j'} {(-1)}^{j+j'+\frac{1}{2}+m_s+m_j+m_j'} \begin{pmatrix} l' & \frac{1}{2} & j' \\ 0 & m_s & -m_j'
\end{pmatrix} \nonumber \\ && \begin{pmatrix} j & L & j' \\ -m_j & 0 & m_j'
\end{pmatrix}
 \begin{pmatrix} j & \frac{1}{2} & l \\ m_j & -m_s & 0
\end{pmatrix} 
\end{eqnarray} 
3j-symbol is non-zero only if  $m_j=m_j'=m_s$
\begin{eqnarray} 
\begin{Bmatrix} l & j & \frac{1}{2} \\ j' & l' & L 
\end{Bmatrix} \begin{pmatrix} l' & L & l \\ 0 & 0 & 0
\end{pmatrix} &=& \sum_{m_s} {(-1)}^{j+j'+\frac{1}{2}+m_s+m_s+m_s} \begin{pmatrix} l' & \frac{1}{2} & j' \\ 0 & m_s & -m_s'
\end{pmatrix} \nonumber \\ &&  \begin{pmatrix} j & L & j' \\ -m_s & 0 & m_s'
\end{pmatrix}
  \begin{pmatrix} j & \frac{1}{2} & l \\ m_s & -m_s & 0
\end{pmatrix} \nonumber 
\end{eqnarray}
\begin{eqnarray}
&=& {(-1)}^{j+j'+\frac{1}{2}+3/2} \begin{pmatrix} l' & \frac{1}{2} & j' \\ 0 & \frac{1}{2} & -\frac{1}{2}
\end{pmatrix} \begin{pmatrix} j & L & j' \\ -\frac{1}{2} & 0 & \frac{1}{2}
\end{pmatrix} \begin{pmatrix} j & \frac{1}{2} & l \\ \frac{1}{2} & -\frac{1}{2} & 0
\end{pmatrix} \nonumber \\
&& + {(-1)}^{j+j'+\frac{1}{2}-3/2}  \begin{pmatrix} l' & \frac{1}{2} & j' \\ 0 & -\frac{1}{2} & \frac{1}{2}
\end{pmatrix} \begin{pmatrix} j & L & j' \\ \frac{1}{2} & 0 & -\frac{1}{2}
\end{pmatrix}
  \begin{pmatrix} j & \frac{1}{2} & l \\ -\frac{1}{2} & \frac{1}{2} & 0
\end{pmatrix} 
\end{eqnarray}
Since, 
$ \begin{pmatrix} j_1 & j_2 & j_3 \\ m_1 & m_2 & m_3 
\end{pmatrix} = {(-1)}^{j_1+j_2+j_3} \begin{pmatrix} j_1 & j_2 & j_3 \\ -m_1 & -m_2 & -m_3 
\end{pmatrix} $,

\begin{eqnarray}
\begin{Bmatrix} l & j & \frac{1}{2} \\ j' & l' & L 
\end{Bmatrix} \begin{pmatrix} l' & L & l \\ 0 & 0 & 0
\end{pmatrix} &=& [{(-1)}^{j+j'+\frac{1}{2}+3/2} \nonumber \\ &&
+{(-1)}^{j+j'+\frac{1}{2}-3/2} {(-1)}^{l'+\frac{1}{2}+j'+j+L+j'+j+\frac{1}{2}+l}] \nonumber \\ && 
\begin{pmatrix} l' & \frac{1}{2} & j' \\ 0 & \frac{1}{2} & -\frac{1}{2}
\end{pmatrix} \begin{pmatrix} j & L & j' \\ -\frac{1}{2} & 0 & \frac{1}{2}
\end{pmatrix}
  \begin{pmatrix} j & \frac{1}{2} & l \\ \frac{1}{2} & -\frac{1}{2} & 0
\end{pmatrix} \nonumber \\
&=& [{(-1)}^{j+j'}+{(-1)}^{j+j'-1} {(-1)}^{l+l'+1+2j+2j'+L}] \nonumber \\ && \begin{pmatrix} l' & \frac{1}{2} & j' \\ 0 & \frac{1}{2} & -\frac{1}{2}
\end{pmatrix} \begin{pmatrix} j & L & j' \\ -\frac{1}{2} & 0 & \frac{1}{2}
\end{pmatrix}
   \begin{pmatrix} j & \frac{1}{2} & l \\ \frac{1}{2} & -\frac{1}{2} & 0
\end{pmatrix} \nonumber \\
&=& {(-1)}^{j+j'} [1+ {(-1)}^{l+l'+1+L}] \nonumber \\ && \begin{pmatrix} l' & \frac{1}{2} & j' \\ 0 & \frac{1}{2} & -\frac{1}{2}
\end{pmatrix} \begin{pmatrix} j & L & j' \\ -\frac{1}{2} & 0 & \frac{1}{2}
\end{pmatrix}  \begin{pmatrix} j & \frac{1}{2} & l \\ \frac{1}{2} & -\frac{1}{2} & 0
\end{pmatrix} \nonumber \\
&=& {(-1)}^{j+j'} [1+ {(-1)}^{l+l'+1+L}] \nonumber \\ && \frac{{(-1)}^{j'-\frac{1}{2}}}{\hat{l'}\sqrt{2}} \begin{pmatrix} j & L & j' \\ -\frac{1}{2} & 0 & \frac{1}{2}
\end{pmatrix} \frac{{(-1)}^{j-\frac{1}{2}}}{\hat{l}\sqrt{2}} \label{103} 
\end{eqnarray} 
Since, 
\begin{eqnarray}
\begin{pmatrix} l' & \frac{1}{2} & j' \\ 0 & \frac{1}{2} & -\frac{1}{2}
\end{pmatrix} &=&  \begin{pmatrix} j' & \frac{1}{2} & l' \\ -\frac{1}{2} & \frac{1}{2} & 0
\end{pmatrix} \nonumber \\
&=&  \frac{{(-1)}^{j'-\frac{1}{2}}}{\hat{l'}\sqrt{2}} \langle j' \quad -\frac{1}{2} \quad  \frac{1}{2} \quad \frac{1}{2} | l' \quad 0 \rangle \nonumber 
\end{eqnarray}
\begin{eqnarray}
\begin{pmatrix} j' & \frac{1}{2} & l' \\ -\frac{1}{2} & \frac{1}{2} & 0
\end{pmatrix} = \frac{{(-1)}^{j'-\frac{1}{2}}}{\hat{l'}\sqrt{2}} \frac{1}{\sqrt{2}} \label{104} \\
\text{and} \begin{pmatrix} j & \frac{1}{2} & l \\ \frac{1}{2} & -\frac{1}{2} & 0
\end{pmatrix} = \frac{{(-1)}^{j-\frac{1}{2}}}{\hat{l}\sqrt{2}} \frac{1}{\sqrt{2}} \label{105}
\end{eqnarray}
From Eqs.(\ref{103}), (\ref{104}) and (\ref{105}), 
\begin{eqnarray}
\begin{Bmatrix} l & j & \frac{1}{2} \\ j' & l' & L 
\end{Bmatrix} \begin{pmatrix} l' & L & l \\ 0 & 0 & 0
\end{pmatrix} &=& {(-1)}^{2j+2j'-1} [1+ {(-1)}^{l+l'+L}] \frac{1}{2\hat{l} \hat{l'}} \nonumber \\ && \begin{pmatrix} j & L & j' \\ -\frac{1}{2} & 0 & \frac{1}{2} \end{pmatrix} \nonumber \\
&=& \frac{-1}{2\hat{l} \hat{l'}} [1+ {(-1)}^{l+l'+L}] \begin{pmatrix} j & L & j' \\ -\frac{1}{2} & 0 & \frac{1}{2} \end{pmatrix} \nonumber \\
&=& \frac{-1}{2\hat{l} \hat{l'}} [1+ {(-1)}^{l+l'+L}] \begin{pmatrix} j' & L & j \\ -\frac{1}{2} & 0 & \frac{1}{2} \end{pmatrix} 
\quad \quad
\end{eqnarray}
Since, $ \begin{pmatrix} j & L & j' \\ -\frac{1}{2} & 0 & \frac{1}{2} \end{pmatrix} = {(-1)}^{j+L+j'} \begin{pmatrix} j & L & j' \\ \frac{1}{2} & 0 & -\frac{1}{2} \end{pmatrix} =  {(-1)}^{2j+2L+2j'} \begin{pmatrix} j' & L & j \\ -\frac{1}{2} & 0 & \frac{1}{2} \end{pmatrix}$. Therefore, the reduced matrix element of spherical harmonics in $| l j>$-basis can be written as
\begin{eqnarray}
\langle l j || Y^L || l' j' \rangle &=& \langle \frac{1}{2} l j || Y^L || \frac{1}{2}l' j' \rangle \nonumber \\
&=& {(-1)}^{\frac{1}{2}+l+l'+j+L} \hat{j} \hat{j'} \hat{l} \hat{l'} \hat{L} \sqrt{\frac{1}{4\pi}} \nonumber \\ && \times \frac{-1}{2 \hat{l} \hat{l'}} [1+ {(-1)}^{l+l'+L}] \begin{pmatrix} j & L & j' \\ -\frac{1}{2} & 0 & \frac{1}{2} \end{pmatrix} \nonumber \\
&=& \frac{-1}{2} { (-1)}^{\frac{1}{2}+l+l'+j+L} [1+ {(-1)}^{l+l'+L}]  \nonumber \\ && \hat{j} \hat{j'} \hat{L}
\sqrt{\frac{1}{4\pi}}  \begin{pmatrix} j & L & j' \\ -\frac{1}{2} & 0 & \frac{1}{2} \end{pmatrix} \nonumber \\
&=& \frac{-1}{2} { (-1)}^{j+\frac{1}{2}} [1+ {(-1)}^{l+l'+L}]  \nonumber \\ && \hat{j} \hat{j'} \hat{L} \sqrt{\frac{1}{4\pi}} \begin{pmatrix} j & L & j' \\ -\frac{1}{2} & 0 & \frac{1}{2} \end{pmatrix} \nonumber \\
&=& \frac{1}{2} { (-1)}^{j+3/2} [1+ {(-1)}^{l+l'+L}]  \nonumber \\ && \hat{j} \hat{j'} \hat{L} \sqrt{\frac{1}{4\pi}} \begin{pmatrix} j & L & j' \\ -\frac{1}{2} & 0 & \frac{1}{2} \end{pmatrix} \nonumber \\
&=& \frac{1}{2} { (-1)}^{j+2-\frac{1}{2}} [1+ {(-1)}^{l+l'+L}]  \nonumber \\ && \hat{j} \hat{j'} \hat{L} \sqrt{\frac{1}{4\pi}} \begin{pmatrix} j & L & j' \\ -\frac{1}{2} & 0 & \frac{1}{2} \end{pmatrix} \nonumber \\
&=& \frac{1}{2} { (-1)}^{j-\frac{1}{2}} [1+ {(-1)}^{l+l'+L}]  \nonumber \\ && \hat{j} \hat{j'} \hat{L} \sqrt{\frac{1}{4\pi}} \begin{pmatrix} j & L & j' \\ -\frac{1}{2} & 0 & \frac{1}{2} \end{pmatrix} 
\end{eqnarray}
If ${(-1)}^{l+l'+L}$ is odd then the reduced matrix element results in zero. On the other hand, if ${(-1)}^{l+l'+L}$ is even then
\begin{eqnarray}
\langle l j || Y^L || l' j' \rangle &=& { (-1)}^{j-\frac{1}{2}} \hat{j} \hat{j'} \hat{L}   \sqrt{\frac{1}{4\pi}} \begin{pmatrix} j & L & j' \\ -\frac{1}{2} & 0 & \frac{1}{2} \end{pmatrix} \quad \quad
\end{eqnarray}
The final result is independent of $l$ and $l'$.

\end{appendices}

\end{document}